\documentclass[aps,prl,10pt,preprint,nofootinbib]{revtex4-1}
\usepackage{amsmath}
\usepackage{amssymb}
\usepackage{color}
\usepackage{gensymb}
\usepackage{textcomp}
\usepackage{graphicx}
\DeclareUnicodeCharacter{0306}{}

\usepackage{epstopdf}
\usepackage{}

\begin{document}

\title{High Harmonic Generation in Solids: Particle and Wave Perspectives}
\author{Liang Li$^{1}$}
\author{Pengfei Lan$^1$}
\email{Corresponding author: pengfeilan@hust.edu.cn}

\author{Xiaosong Zhu$^1$}
\author{Peixiang Lu$^{1,2,3}$}\email{Corresponding author: lupeixiang@hust.edu.cn}

\affiliation{
	$^1$Wuhan National Laboratory for Optoelectronics and School of Physics, Huazhong University of Science and Technology, Wuhan 430074, China\\
	$^2$Hubei Key Laboratory of Optical Information and Pattern Recognition, Wuhan Institute of Technology, Wuhan 430205, China\\
	$^3$CAS Center for Excellence in Ultra-intense Laser Science, Shanghai 201800, China
}


\begin{abstract}
High harmonic generation (HHG) from gas phase atoms (or molecules) has opened up a new frontier in ultrafast optics, where attosecond time resolution and Angstrom spatial resolution are accessible. The fundamental physical pictures of HHG is always explained by the laser-induced recollision of particle-like electron motion, which lay the foundation of attosecond spectroscopy. In recent years, HHG has also been observed in solids. One can expect the extension of attosecond spectroscopy to the condensed matter if a description capable of resolving ultrafast dynamics is provided. Thus, a large number of theoretical studies have been proposed to understand the underlying physics of HHG. Here, we revisit the recollision picture in solid HHG and show some challenges of current methods with particle perspective, and present the recently developed wave perspective Huygens-Fresnel picture in understanding dynamical systems within the ambit of strong-field physics.
\end{abstract}

\maketitle
\section{1. Introduction}
The last three decades have seen the breakthrough of ultrafast science due to the generation of high-order harmonics \cite{mcpherson1987,ferray1988,krause1992,corkum1993,schafer1993,macklin1993} and attosecond pulse \cite{hentschel2001,paul2001,sansone2006} by using a near-infrared femtosecond laser. Just after the observation of high harmonic generation (HHG), the plateau
structure of the spectra, which can not be explained by traditional perturbation theory,
has attracted lots of interest, and the pioneer studies eventually led to great achievements:
“three-step” recollision model \cite{krause1992,corkum1993}. When an atom or molecule is exposed to an intense short infrared laser pulse, an electron that was ejected at an earlier time may be driven back by the oscillating electric field to revisit its parent ion. When the electron recollides with its parent ion, recombination, and rescattering take place. Recombination results in the emission of coherent XUV radiation, i.e., HHG \cite{itatani2004,blaga2012,pullen2015,sekikawa2004,gohle2005,nabekawa2006}. Rescattering can result in above-threshold ionization (ATI) \cite{agostini1979,javanainen1988,lewenstein1995} and nonsequential double ionization (NSDI) \cite{fittinghoff1992,walker1994,kondo1993}. An analytic model called strong field approximation (SFA) \cite{lewenstein1994} was established following the idea of “three-step” recollision. A nice review of the SFA model is referred to Ref. \cite{Amini2019}. The “three-step” recollision model provides an intuitive basis for understanding the electron dynamics underlying the strong-field processes \cite{chen2000,wu2013,kamor2014,wei2017,milovsevic2017}. It is of great importance and value to attosecond control, for example, HHG can be controlled by controlling the “three-step” recollision process so as to produce isolated attosecond pulse \cite{hentschel2001,sansone2006,Pfeifer2006,Lan2007,Mashiko_2008,Takahashi_2013}. Moreover, the recollision itself encodes rich information about the atomic or molecular structure and electron dynamics. Within this process, a moment of ionization maps into a specific time and energy of recollision, and an electron interferometer, created by light with producing high harmonics or photoelectrons is established. It reveals the self-detection behavior of coherent electrons induced by strong lasers, which has stimulated advanced attosecond spectroscopy methods e.g., high harmonic spectroscopy (HHS) \cite{baker2006,smirnova2009,haessler2010,vozzi2011,kraus2015,lan2017,Bruner2015,Camper2023,He2022,Tuthill2020} and laser-induced electron diffraction (LIED) \cite{zuo1996,yurchenko2004,morishita2008,xu2012,wolter2016}, for imaging and steering the electron dynamics with unprecedented attosecond-sub-angstrom ultrahigh spatiotemporal resolution.

In the last decade, HHG has also been observed in solids \cite{ghimire2011,luu2015,vampa2015,langer2016,banks2017,von2018,jiang2018,uzan2020}, which makes it possible to extend the successful attosecond science to condensed-matter systems. Comprehensive knowledge and control of the electron dynamics in condensed-matter systems are pertinent to the development of many modern technologies, such as petahertz electronics \cite{krausz2014,luu2015,schotz2019,sederberg2020}, optoelectronics \cite{Gu_2012,schiffrin2013,schultze2013,vampa2018}, information processing \cite{hohenleutner2015,luu2015,langer2017,higuchi2017,Chakraborty_2015}, and photovoltaics \cite{falke2014,andrea2013,Cook_2017}. Therefore, there is a growing interest to extend the attosecond science to solid systems. However, the understanding of the underlying physics is limited by the complicated structure and dynamical processes in solids and therefore the extension of the well-developed attosecond techniques from gas to solids still faces lots of challenges. Although several numerical models, such as time-dependent Schödinger equation (TDSE) \cite{wu2015,wu2017,liu2017,du2017,ikemachi2017,Plaja_1992,Huang2017}, semiconductor Bloch equations (SBEs) \cite{golde2008,haug2009,kira2011,vampa2014,F_ldi_2013,McDonald2015,Tamaya2016} and time-dependent density functional theory (TDDFT) \cite{runge1984,tancogne2017,tancogne2018,bauer2018,neufeld2021,Hansen_2017,Hansen_2018} can give good descriptions of HHG, the underlying mechanisms is buried in the wave functions. In order not to conceal the physics behind mathematics, two main mechanisms are usually considered to explain the solid HHG: interband polarization and intraband current. The former mechanism has much in common with the ``three-step'' recollision model of gas HHG. The laser promotes an electron from the valence band to the conduction band, leaving a hole in the VB, then the electron is accelerated by the laser field, and finally, high harmonics can be produced when the electron recombines (or rescatters) with the hole. The intraband mechanism also starts from the electron excitation from the valence band to the conduction band. Then the electron and hole move anharmonically in the strong laser field, which leads to an anharmonic current and high harmonics. Different from the interband polarization, the motion of the electron and hole in the intraband current mechanism are separated. Which mechanism dominates the solid HHG depends on the band gap of the solid and also the driving laser pulse \cite{Ortmann_2021}. Generally, in a short-wave infrared laser, the interband mechanism is dominant for the high harmonic above the band gap, i.e., in the plateau of the spectra. In contrast, the intraband mechanism dominates at longer driving wavelength or the high harmonics below the band gap. Due to the similarity between the interband mechanism and the ``three step'' model, the interband mechanism has attracted a lot of attention and it is also in favor of constructing the HHS techniques in solids. Therefore, some generalized recollision models \cite{vampa2015semiclassical,osika2017,yue2020,parks2020} are proposed following the counterpart of HHG in gas phase. Within the frame of these models, the electronic response under the influence of intense laser fields leads to a multi-dimensional integral, which comes from the expression of the currents (time-dependent dipole momentum) within a factorial form under SFA. Then, the saddle point approximation is applied considering the form of highly oscillatory integrals, and the recollision picture of electron dynamics is derived by the saddle point equations. These models provide useful and intuitive explanations of HHG in solids, which established a connection between attosecond physics in gas and condensed-matter phases, and promoted the development of HHS in solids \cite{Uzan_Narovlansky_2022,vampa2015,hohenleutner2015,langer2016,You2016,Luu_2018,Vampa_2015,Uchida_2021}.

As is well known, wave-particle duality is the elementary nature of the quantum process. The electron is much less localized in solids compared to that in gases. Moreover, solid systems always have complicated band structures and thus the diffraction of electron wave packets may become prominent. Very recently, a Huygens-Fresnel picture for solid HHG is proposed \cite{li2021}. In this picture, the electron motion is described by a series of wavelets instead of classical particles, and the HHG is interpreted as a coherent superposition of all wavelet contributions. This stimulates a different paradigm and idea within the rapidly emerging theoretical studies for ultrafast electrodynamic processes in strong fields.

In the last few years, several reviews have been published that systematically summarized the present numerical and theoretical methods in the development of HHG in solids \cite{yu2019,yue2022,Ortmann_2021,Wang_2022,Goulielmakis_2022,Park_2021,Kruchinin_2018}. In this review, we focus on the semiclassical perspectives of solid HHG which aims to decode the phase information of coherent electron dynamics, highlighting the commonality and differences of the representative models. We revisit the recollision picture for high-harmonic generation in solids, which lies at the heart of HHG and attosecond science, and present a perspective by treating the electron as a particle and wave packet. The rest of this review is organized as follows. In Section 2, we summarize the present recollision models and show some challenges and special issues to be improved or solved. In Section 3, we introduce the recently proposed Huygens-Fresnel picture for HHG in solids. In Section 4, the differences and unifications between particle-like recollision picture and wave-like Huygens-Fresnel picture are clarified, and the advantages and drawbacks of the different methods are compared. Moreover, we present a concrete calculation example to exemplify how electronic volatility acts on the HHG process. In Section 5, we present a brief prospect of the wave-like picture in attosecond science.

\section{2. Recollision models for HHG in solids}

HHG in solids can result from interband and intraband mechanisms \cite{ghimire2011,Ghimire2012,Higuchi2014,You2016}. For the commonly used short wave infrared laser, the interband current is usually dominant as demonstrated by previous experiments \cite{vampa2015,Kaneshima2018,Li_2021} and theoretical results \cite{vampa2015semiclassical,vampa2014}. In a similar fashion as atomic HHG, saddle point equations have also been applied to the semiclassical interpretation of interband HHG, which stimulates many theoretical models \cite{vampa2014,vampa2015semiclassical,osika2017,yue2020,parks2020}, e.g., the classical recollision model, non-perfect recollision model and so on. Here, we briefly review these recollision models.

\subsection{A. Derivation of HHG in solids}
We start with a theoretical derivation of the HHG in solids. Although these descriptions have been discussed in recent reviews \cite{yu2019,yue2022}, we still present the relevant results for completeness and for the convenience of further discussions. Note that the aim of this review is to elucidate one of the numerous aspects of HHG in solid, namely the recollision picture. Some phenomena, such as multi-electron effects and propagation effects, are out of this paper. Therefore, here we only show the derivation of HHG from a single-electron Hamiltonian.

The HHG process in solids is modeled by a nonrelativistic electron in a periodic potential interacting with an external electromagnetic field, which is described by a minimal-coupling Hamiltonian \cite{Scully1997a} (atomic units (a.u.) are applied in this work unless stated)
\begin{align}\label{Hamiltonian}
\hat{H}(t) = \frac{[\hat{\textbf{p}}+\textbf{A(\textbf{r},t)}]^2}{2}-U(\textbf{r},t)+V(\textbf{r}),
\end{align}
where $\textbf{A}(\textbf{r},t)$ and $U(\textbf{r},t)$ are the vector and scalar potentials of the external field, respectively, and $V(\textbf{r})$ is the crystal periodic potential. There is a gauge freedom in choosing $\textbf{A}(\textbf{r},t)$ and $U(\textbf{r},t)$,
\begin{align}\label{Gauge Transformation}
\textbf{A}(\textbf{r},t) \rightarrow \textbf{A}(\textbf{r},t)+\nabla\chi(\textbf{r},t) \nonumber\\
U(\textbf{r},t) \rightarrow U(\textbf{r},t)-\partial_{t}\chi(\textbf{r},t),
\end{align}
with $\chi(\textbf{r},t)$ a differentiable real function. The gauge-independent quantities are the electric and magnetic fields
\begin{align}\label{electromagnetic field}
\textbf{F} &= -\nabla U-\partial_{t}\textbf{A}, \\
\textbf{B} &= \nabla \times \textbf{A}.
\end{align}
The minimal-coupling Hamiltonian (\ref{Hamiltonian}) can be reduced to a simple form by using the dipole approximation considering that the wavelengths of driving fields used for HHG are much larger than the dimension of the unit cell. In this case, the vector potential can be written in dipole approximation, $\textbf{k}\cdot\textbf{r}\ll1$, as $\textbf{A}(t) \equiv \textbf{A}(\textbf{r},t)$. The TDSE for this problem (in the dipole approximation) is given by
\begin{align}\label{TDSE}
i\partial_{t}\lvert\Psi(t)\rangle = H_{\text{VG}}(t)\lvert\Psi(t)\rangle =  \left[\frac{[\hat{\textbf{p}}+\textbf{A}(t)]^2}{2}+V(\textbf{r})\right]\lvert\Psi(t)\rangle.
\end{align}
Here, we are applying the velocity gauge, in which $U(\textbf{r},t) = 0$ and $\textbf{A}(t) =-\int_{-\infty}^{t}\textbf{F}(\tau)\mathrm{d}\tau$.

We first consider the field-free states before the laser field is turned on. The eigenvalue equation can be written as
\begin{align}\label{eigenvalue equation}
\left[\frac{\hat{\textbf{p}^{2}}}{2}+V(\textbf{r})\right]\psi_{i,\textbf{k}}(\textbf{r}) = E_{i}(\textbf{k})\psi_{i,\textbf{k}}(\textbf{r}).
\end{align}
In this equation, $\psi_{i,\textbf{k}}(\textbf{r})$ and $E_{i}(\textbf{k})$ are the Bloch function and the energy band of the field-free crystal with a band index $i$ and a crystal momentum $\textbf{k}$. In the coordinate representation, $\psi_{i,\textbf{k}}(\textbf{r})$ is a product of a plane wave and a periodic envelope function
\begin{align}\label{Bloch function}
\psi_{i,\textbf{k}}(\textbf{r}) = e^{i\textbf{k}\cdot\textbf{r}}u_{i,\textbf{k}}(\textbf{r}),
\end{align}
where $u_{i,\textbf{k}}(\textbf{r+R})=u_{i,\textbf{k}}(\textbf{r})$ for all $\textbf{R}$ from the Bravais lattice.

Let us now consider the instantaneous eigenstates of $H_{\text{VG}}(t)$ in the presence of a homogeneous external field, which satisfies
\begin{align}\label{instantaneous eigenstates}
\left(\frac{[\hat{\textbf{p}}+\textbf{A}(t)]^2}{2}+V(\textbf{r})\right)|\varphi_{i,\textbf{k}_{0}}(t)\rangle = \widetilde{E}_{i,\textbf{k}_{0}}(t)|\varphi_{i,\textbf{k}_{0}}(t)\rangle.
\end{align}
Since the Hamiltonian is periodic in space, the Bloch theorem is applicable. The solution of Eq. (\ref{instantaneous eigenstates}) satisfying the Born-von K\'{a}rm\'{a}n boundary condition yields \cite{Krieger_1986}
\begin{align}\label{H_state}
\langle r\lvert\varphi_{i,\textbf{k}_{0}}(t)\rangle &= e^{-i\textbf{A}(t)\cdot\textbf{r}}\psi_{i,\textbf{k}(t)}(\textbf{r}),   \\
\widetilde{E}_{i,\textbf{k}_{0}}(t) &= E_{i}\textbf{(}\textbf{k}(t)\textbf{)}.
\end{align}
The states $\lvert\varphi_{i,\textbf{k}_{0}}(t)\rangle$ are called accelerated Bloch states or Houston functions, and the time-dependent crystal momentum $\textbf{k}(t) = \textbf{k}_{0}+\textbf{A}(t)$ satisfies the acceleration theorem: $\frac{\mathrm{d}\textbf{k}}{\mathrm{d}t}=-\textbf{F}(t)$.

Then, we use the Houston functions as a basis for solving the TDSE with the ansatz
\begin{align}\label{wave function}
\lvert\Psi_{\textbf{k}_{0}}(t)\rangle &= \sum_{m}\alpha_{m,\textbf{k}_{0}}(t)\lvert\varphi_{m,\textbf{k}_{0}}(t)\rangle.
\end{align}
By inserting this ansatz into TDSE and projecting onto $\langle\varphi_{n,\textbf{k}_{0}}(t)|$, one gets
\begin{align}\label{TDSE_with_ansatz}
&i\partial_{t}\alpha_{n,\textbf{k}_{0}}(t)+\sum_{m}\alpha_{m,\textbf{k}_{0}}(t)\langle\varphi_{n,\textbf{k}_{0}}(t)|i\partial_{t}\lvert\varphi_{m,\textbf{k}_{0}}(t)\rangle  \nonumber\\
&= \langle\varphi_{n,\textbf{k}_{0}}(t)|H_{VG}(t)\left[\sum_{m}\alpha_{m,\textbf{k}_{0}}(t)\lvert\varphi_{m,\textbf{k}_{0}}(t)\rangle\right] = E_{n}\textbf{(}\textbf{k}(t)\textbf{)}\alpha_{n,\textbf{k}_{0}}(t).
\end{align}
Note that $\langle r\lvert\varphi_{m,\textbf{k}_{0}}(t)\rangle = e^{-i\textbf{A}(t)\cdot\textbf{r}}\psi_{m,\textbf{k}(t)}(\textbf{r}) = e^{i\textbf{k}_{0}\cdot\textbf{r}}u_{m,\textbf{k}(t)}(\textbf{r})$ and the nonadiabatic couplings are $\langle\varphi_{n,\textbf{k}_{0}}(t)|i\partial_{t}\lvert\varphi_{m,\textbf{k}_{0}}(t)\rangle = -\textbf{F}(t)\cdot\langle u_{n,\textbf{k}(t)}|i\nabla_{\textbf{k}}|u_{m,\textbf{k}(t)}\rangle = -\textbf{F}(t)\cdot\textbf{d}_{nm}\textbf{(}\textbf{k}(t)\textbf{)}$. Then, one can obtain the following system of differential equations
\begin{align}\label{TDSE_with_Houston_basis}
i\partial_{t}\alpha_{n,\textbf{k}_{0}}(t) = E_{n}\textbf{(}\textbf{k}(t)\textbf{)}\alpha_{n,\textbf{k}_{0}}(t)+\textbf{F}(t)\cdot\sum_{l}\textbf{d}_{ml}\textbf{(}\textbf{k}(t)\textbf{)}\alpha_{l,\textbf{k}_{0}}(t).
\end{align}

The connection between the TDSE and SBEs can be established by introducing the density matrix elements $\rho_{mn}^{\textbf{k}_{0}} = \alpha_{m,\textbf{k}_{0}}\alpha_{n,\textbf{k}_{0}}^{*}$. Multiplying Eq. (\ref{TDSE_with_Houston_basis}) by $\alpha_{n,\textbf{k}_{0}}^{*}$ and utilizing the complex conjugate of the resulting equations, Eqs. (\ref{TDSE_with_Houston_basis}) can be rewritten into differential equations
\begin{align}\label{density-matrix-equations}
i\partial_{t}\rho_{mn}^{\textbf{k}_{0}}(t) = [E_{m}\textbf{(}\textbf{k}(t)\textbf{)}-E_{n}\textbf{(}\textbf{k}(t)\textbf{)}]\rho_{mn}^{\textbf{k}_{0}}(t)+\textbf{F}(t)\cdot\sum_{l}\left[\textbf{d}_{ml}\textbf{(}\textbf{k}(t)\textbf{)}\rho_{ln}^{\textbf{k}_{0}}(t)-\rho_{ml}^{\textbf{k}_{0}}(t)\textbf{d}_{ln}\textbf{(}\textbf{k}(t)\textbf{)}\right].
\end{align}
A dephasing term $-i(1-\delta_{mn})\rho_{mn}^{\textbf{k}_{0}(t)}/T_{2}$ can be introduced on the right-hand side of Eqs. (\ref{density-matrix-equations}), which phenomenologically describes the many-body couplings such as electron–electron and electron–phonon scattering.

Finally, HHG in solids is determined by the microscopic current $\textbf{J}(t) = \sum_{\textbf{k}_0}\textbf{j}_{\textbf{k}_0}(t)$, where the contribution from an electron with an initial crystal momentum $\textbf{k}_{0}$ is evaluated as
\begin{align}\label{microscopic current}
\textbf{j}_{\textbf{k}_0}(t) &= -\langle\Psi_{\textbf{k}_0}(t)|\left[\hat{\textbf{p}}+\textbf{A}(t)\right]|\Psi_{\textbf{k}_0}(t)\rangle \nonumber\\
&= -\sum_{m,n}\langle\Psi_{\textbf{k}_0}(t)|\varphi_{m,\textbf{k}_{0}}(t)\rangle\langle\varphi_{m,\textbf{k}_{0}}(t)|\left[\hat{\textbf{p}}+\textbf{A}(t)\right]|\varphi_{n,\textbf{k}_{0}}(t)\rangle\langle\varphi_{n,\textbf{k}_{0}}(t)|\Psi_{\textbf{k}_0}(t)\rangle  \nonumber\\
&= -\sum_{m,n}\langle\varphi_{m,\textbf{k}_{0}}(t)|\left[\hat{\textbf{p}}+\textbf{A}(t)\right]|\varphi_{n,\textbf{k}_{0}}(t)\rangle\rho^{\textbf{k}_{0}}_{nm}(t) \nonumber\\
&= -\sum_{m,n}\textbf{p}_{mn}^{\textbf{k}_{0}+\textbf{A}(t)}\rho^{\textbf{k}_{0}}_{nm}(t).
\end{align}
The momentum matrix elements can be determined by using the relation $\left[\hat{\textbf{p}}+\textbf{A}(t)\right] = -i\left[\hat{\textbf{r}},\hat{H}_{VG}(t)\right]$, that is
\begin{align}\label{momentum matrix elements}
\textbf{p}_{mn}^{\textbf{k}_{0}+\textbf{A}(t)} &= \langle\varphi_{m,\textbf{k}_{0}}(t)|\left[\hat{\textbf{p}}+\textbf{A}(t)\right]|\varphi_{n,\textbf{k}_{0}}(t)\rangle   \nonumber\\
&= -i\left\langle\varphi_{m,\textbf{k}_{0}}(t)\left|\left[\hat{\textbf{r}},\hat{H}_{VG}(t)\right]\right|\varphi_{n,\textbf{k}_{0}}(t)\right\rangle.
\end{align}
It is convenient to split the position operator into an intraband part $\hat{\textbf{r}}_{i}$ and an interband part $\hat{\textbf{r}}_{e}$ \cite{Aversa_1995}, where $\hat{\textbf{r}} = \hat{\textbf{r}}_{i}+\hat{\textbf{r}}_{e}$ and
\begin{align}\label{position matrix elements}
\langle\psi_{m,\textbf{k}}|\hat{\textbf{r}}_{i}|\psi_{n,\textbf{k}'}\rangle &= \delta_{mn}\left[\delta(\textbf{k}-\textbf{k}')\textbf{d}_{mm}({\textbf{k}})+i\nabla_{\textbf{k}}\delta(\textbf{k}-\textbf{k}')\right],   \\
\langle\psi_{m,\textbf{k}}|\hat{\textbf{r}}_{e}|\psi_{n,\textbf{k}'}\rangle &= \left(1-\delta_{mn}\right)\delta(\textbf{k}-\textbf{k}')\textbf{d}_{mn}({\textbf{k}}).
\end{align}
The relation between $\textbf{d}_{mn}({\textbf{k}})$ and the momentum matrix elements is given by
\begin{align}\label{position and momentum matrix elements}
\textbf{p}_{mn}^{\textbf{k}_{0}+\textbf{A}(t)} &= -i\left\langle\varphi_{m,\textbf{k}_{0}}(t)\left|\left[\hat{\textbf{r}}\hat{H}_{VG}(t)-\hat{H}_{VG}(t)\hat{\textbf{r}}\right]\right|\varphi_{n,\textbf{k}_{0}}(t)\right\rangle   \nonumber\\
&= \sum_{l,\textbf{k}'}-i\langle\varphi_{m,\textbf{k}_{0}}(t)|\hat{\textbf{r}}|\varphi_{l,\textbf{k}'}(t)\rangle
\langle\varphi_{l,\textbf{k}'}(t)|\hat{H}_{VG}(t)|\varphi_{n,\textbf{k}_{0}}(t)\rangle   \nonumber\\
&\ \ \ \ +i\langle\varphi_{m,\textbf{k}_{0}}(t)|\hat{H}_{VG}(t)|\varphi_{l,\textbf{k}'}(t)\rangle
\langle\varphi_{l,\textbf{k}'}(t)|\hat{\textbf{r}}|\varphi_{n,\textbf{k}_{0}}(t)\rangle   \nonumber\\
&= \sum_{l,\textbf{k}'}-i\langle\psi_{m,\textbf{k}_{0}+\textbf{A}(t)}|\hat{\textbf{r}}|\psi_{l,\textbf{k}'+\textbf{A}(t)}\rangle E_{n}\textbf{(}\textbf{k}_{0}+\textbf{A}(t)\textbf{)}\delta_{ln}\delta(\textbf{k}-\textbf{k}')\nonumber\\
&\ \ \ \ +iE_{l}\textbf{(}\textbf{k}'+\textbf{A}(t)\textbf{)}\delta_{ml}\delta(\textbf{k}-\textbf{k}')
\langle\psi_{l,\textbf{k}'+\textbf{A}(t)}|\hat{\textbf{r}}|\psi_{n,\textbf{k}_{0}+\textbf{A}(t)}\rangle   \nonumber\\
&= \delta_{mn}\left[\nabla_{\textbf{k}}E_{n}\textbf{(}\textbf{k}(t)\textbf{)}\right]+i(1-\delta_{mn})\left[E_{m}\textbf{(}\textbf{k}(t)\textbf{)}-E_{n}\textbf{(}\textbf{k}(t)\textbf{)}\right]\textbf{d}_{mn}\textbf{(}\textbf{k}(t)\textbf{)}.
\end{align}
Using the above relations, the microscopic current can be reorganized as an intraband and an interband contribution, i.e., $\textbf{J}(t) = \textbf{J}_{\text{er}}(t)+\textbf{J}_{\text{ra}}(t)$:
\begin{align}\label{inter_and_intra_current}
\textbf{J}_{\text{ra}}(t)&= -\sum_{m,\textbf{k}_{0}}\nabla_{\textbf{k}}E_{m}\textbf{(}\textbf{k}(t)\textbf{)}\rho^{\textbf{k}_{0}}_{mm}(t),   \\
\textbf{J}_{\text{er}}(t)&= -i\sum_{m\neq n,\textbf{k}_{0}}\left[E_{m}\textbf{(}\textbf{k}(t)\textbf{)}-E_{n}\textbf{(}\textbf{k}(t)\textbf{)}\right]\textbf{d}_{mn}\textbf{(}\textbf{k}(t)\textbf{)}\rho^{\textbf{k}_{0}}_{nm}(t).
\end{align}

For simplicity, we only consider an initially fully filled valence band (``$m=v$'') and an empty conduction band (``$m=c$'') in our following discussions. In this case, Eq. (\ref{density-matrix-equations}) evolves as the two-band SBEs
\begin{align}\label{SBEs}
i\partial_{t}\rho_{cv}^{\textbf{k}_{0}}(t) &= \left\{\omega_{g}^{\textbf{k}(t)}+\textbf{F}(t)\cdot\left[\mathbf{\Lambda}_{c}^{\textbf{k}(t)}-\mathbf{\Lambda}_{v}^{\textbf{k}(t)}\right]\right\}\rho_{cv}^{\textbf{k}_{0}}(t)   \nonumber\\
&\ \ \ \  +\textbf{F}(t)\cdot\textbf{d}^{\textbf{k}(t)}\left[\rho_{vv}^{\textbf{k}_{0}}(t)-\rho_{cc}^{\textbf{k}_{0}}(t)\right], \\
i\partial_{t}\rho_{cc}^{\textbf{k}_{0}}(t) &= \textbf{F}(t)\cdot\left\{\textbf{d}^{\textbf{k}(t)}\left[\rho_{cv}^{\textbf{k}_{0}}(t)\right]^{*}-\left[\textbf{d}^{\textbf{k}(t)}\right]^{*}\rho_{cv}^{\textbf{k}_{0}}(t)\right\},   \\
i\partial_{t}\rho_{vv}^{\textbf{k}_{0}}(t) &= -\textbf{F}(t)\cdot\left\{\textbf{d}^{\textbf{k}(t)}\left[\rho_{cv}^{\textbf{k}_{0}}(t)\right]^{*}-\left[\textbf{d}^{\textbf{k}(t)}\right]^{*}\rho_{cv}^{\textbf{k}_{0}}(t)\right\}.
\end{align}
$\omega_{g}^{\textbf{k}} = E_{c}(\textbf{k}) - E_{v}(\textbf{k})$, $\mathbf{\Lambda}_{m}^{\textbf{k}} = \textbf{d}_{mm}(\textbf{k})$ and $\textbf{d}^{\textbf{k}} = \textbf{d}_{cv}(\textbf{k})$ denote the band gap, Berry connection and transition dipole momentum, respectively. The SBEs can be further simplified by a transformation using an integrating factor
\begin{align}\label{transformation}
\alpha_{m,\textbf{k}_{0}}(t) = \tilde{\alpha}_{m,\textbf{k}_{0}}(t)e^{i\phi_{m}^{\text{D}}\textbf{(}\textbf{k}(t)\textbf{)}+i\phi_{m}^{\text{B}}\textbf{(}\textbf{k}(t)\textbf{)}},
\end{align}
where the dynamic phase $\phi_{m}^{\text{D}}\textbf{(}\textbf{k}(t)\textbf{)}$ and the Berry phase $\phi_{m}^{\text{B}}\textbf{(}\textbf{k}(t)\textbf{)}$ are defined by the following expressions
\begin{align}\label{Dynamic and Berry phases}
\phi_{m}^{\text{D}}\textbf{(}\textbf{k}(t)\textbf{)} &= -\int_{-\infty}^{t}E_{m}\textbf{(}\textbf{k}(\tau)\textbf{)}\mathrm{d}\tau,   \\
\phi_{m}^{\text{B}}\textbf{(}\textbf{k}(t)\textbf{)} &= -\int_{-\infty}^{t}\textbf{F}(\tau)\cdot\mathbf{\Lambda}_{m}^{\textbf{k}(\tau)}\mathrm{d}\tau.
\end{align}
Using the notations $\Pi^{\textbf{k}_{0}}(t) = \tilde{\alpha}_{c,\textbf{k}_{0}}\tilde{\alpha}_{v,\textbf{k}_{0}}^{*} = \rho_{cv}^{\textbf{k}_{0}}(t)e^{-i\left[\phi_{cv}^{\text{D}}\textbf{(}\textbf{k}(t)\textbf{)}+\phi_{cv}^{\text{B}}\textbf{(}\textbf{k}(t)\textbf{)}\right]}$, $N_{m}^{\textbf{k}_{0}}(t) = \tilde{\alpha}_{m,\textbf{k}_{0}}\tilde{\alpha}_{m,\textbf{k}_{0}}^{*} = \rho_{mm}^{\textbf{k}_{0}}(t)$ ($m = c,v$) and rearranging Eqs. (22-24), one gets
\begin{align}\label{transformed SBEs}
\partial_{t}\Pi^{\textbf{k}_{0}}(t) &= -i\textbf{F}(t)\cdot\textbf{d}^{\textbf{k}(t)}\left[N_{v}^{\textbf{k}_{0}}(t)-N_{c}^{\textbf{k}_{0}}(t)\right]e^{-i\left[\phi_{cv}^{\text{D}}\textbf{(}\textbf{k}(t)\textbf{)}+\phi_{cv}^{\text{B}}\textbf{(}\textbf{k}(t)\textbf{)}\right]}, \\
\partial_{t}N_{c}^{\textbf{k}_{0}}(t) &=  2\text{Re}\left\{i\textbf{F}(t)\cdot\left[\textbf{d}^{\textbf{k}(t)}\right]^{*}\Pi^{\textbf{k}_{0}}(t)e^{i\left[\phi_{cv}^{\text{D}}\textbf{(}\textbf{k}(t)\textbf{)}+\phi_{cv}^{\text{B}}\textbf{(}\textbf{k}(t)\textbf{)}\right]}\right\},   \\
\partial_{t}N_{v}^{\textbf{k}_{0}}(t) &= - 2\text{Re}\left\{i\textbf{F}(t)\cdot\left[\textbf{d}^{\textbf{k}(t)}\right]^{*}\Pi^{\textbf{k}_{0}}(t)e^{i\left[\phi_{cv}^{\text{D}}\textbf{(}\textbf{k}(t)\textbf{)}+\phi_{cv}^{\text{B}}\textbf{(}\textbf{k}(t)\textbf{)}\right]}\right\}, \label{transformed SBEs 3}
\end{align}
where $\phi_{cv}^{D}\textbf{(}\textbf{k}(t)\textbf{)} = \phi_{c}^{D}\textbf{(}\textbf{k}(t)\textbf{)}-\phi_{v}^{D}\textbf{(}\textbf{k}(t)\textbf{)}$ and $\phi_{cv}^{B}\textbf{(}\textbf{k}(t)\textbf{)} = \phi_{c}^{B}\textbf{(}\textbf{k}(t)\textbf{)}-\phi_{v}^{B}\textbf{(}\textbf{k}(t)\textbf{)}$ are the transitional dynamic phase and Berry phase, respectively. Note that for HHG in semiconductors and insulators, the electrons are initially in the valence band before the external field is turned on (i.e., $N_{v}^{\textbf{k}_{0}}(t_{0}) = 1$) and the population transfer to the conduction band is small. Thus, it is reasonable to explore Eqs. (\ref{transformed SBEs}) by using the Keldysh approximation $N_{v}^{\textbf{k}_{0}}(t)-N_{c}^{\textbf{k}_{0}}(t) \approx 1$. This decouples Eqs. (\ref{transformed SBEs})-(\ref{transformed SBEs 3}) so that they can be formally integrated,
\begin{align}\label{solution for SBEs}
\Pi^{\textbf{k}_{0}}(t) &= -i\int_{t_{0}}^{t}\textbf{F}(t')\cdot\textbf{d}^{\textbf{k}(t')}e^{-i\left[\phi_{cv}^{\text{D}}\textbf{(}\textbf{k}(t')\textbf{)}+\phi_{cv}^{\text{B}}\textbf{(}\textbf{k}(t')\textbf{)}\right]}\mathrm{d}t',   \\
N_{c}^{\textbf{k}_{0}}(t) &= i\int_{t_{0}}^{t}\textbf{F}(t')\cdot\left[\textbf{d}^{\textbf{k}(t')}\right]^{*}\Pi^{\textbf{k}_{0}}(t')e^{i\left[\phi_{cv}^{\text{D}}\textbf{(}\textbf{k}(t')\textbf{)}+\phi_{cv}^{\text{B}}\textbf{(}\textbf{k}(t')\textbf{)}\right]}\mathrm{d}t'+\text{c.c.},   \\
N_{v}^{\textbf{k}_{0}}(t) &= N_{v}^{\textbf{k}_{0}}(t_{0})-N_{c}^{\textbf{k}_{0}}(t).
\end{align}
Inserting the results into Eqs. (20) and (21), we find
\begin{align}\label{integrated currents}
\textbf{J}_{\text{ra}}(t) &= -\sum_{\textbf{k}_0}\nabla_{\textbf{k}}\omega_{g}^{\textbf{k}(t)}\int_{t_{0}}^{t}\mathrm{d}t'\textbf{F}(t')\cdot\left[\textbf{d}^{\textbf{k}(t')}\right]^{*}e^{i\left[\phi_{cv}^{\text{D}}\textbf{(}\textbf{k}(t')\textbf{)}+\phi_{cv}^{\text{B}}\textbf{(}\textbf{k}(t')\textbf{)}\right]}   \nonumber\\
&\ \ \ \ \times \int_{t_{0}}^{t'}\textbf{F}(t'')\cdot\textbf{d}^{\textbf{k}(t'')}e^{-i\left[\phi_{cv}^{\text{D}}\textbf{(}\textbf{k}(t'')\textbf{)}+\phi_{cv}^{\text{B}}\textbf{(}\textbf{k}(t'')\textbf{)}\right]}\mathrm{d}t''+\text{c.c.},   \\
\textbf{J}_{\text{er}}(t)&= \sum_{\textbf{k}_0}\omega_{g}^{\textbf{k}(t)}\left[\textbf{d}^{\textbf{k}(t)}\right]^{*}e^{i\left[\phi_{cv}^{\text{D}}\textbf{(}\textbf{k}(t)\textbf{)}+\phi_{cv}^{\text{B}}\textbf{(}\textbf{k}(t)\textbf{)}\right]}\int_{t_{0}}^{t}\textbf{F}(t')\cdot\textbf{d}^{\textbf{k}(t')}e^{-i\left[\phi_{cv}^{\text{D}}\textbf{(}\textbf{k}(t')\textbf{)}+\phi_{cv}^{\text{B}}\textbf{(}\textbf{k}(t')\textbf{)}\right]}\mathrm{d}t'+\text{c.c.}.
\end{align}
The HHG spectrum can be obtained by Fourier transformation of the currents.

\subsection{B. Saddle point equations and the recollision models}
In the above, we have shown the derivation of HHG in solids. The relevant observables, i.e., the induced currents, are expressed as an integral form in the multi-dimensional space by using the Keldysh approximation. These integrals can often be solved using the stationary phase approximation, which leads to a series of equations identifying the points in the multi-dimensional space having the most significant contributions in their evaluation.
These points are usually indicated as saddle points. Such an approach enables an approximate intuitive physical picture with classical perceptions of the quantum mechanical processes under investigation. Thus, the saddle point methods are very powerful and valuable general theoretical tools to obtain asymptotic expressions of mathematical solutions and also help to gain physical insights into the underlying phenomena. Such techniques have been adapted to study the attosecond science in gaseous systems in the past \cite{baker2006,smirnova2009,haessler2010,vozzi2011,kraus2015,lan2017,yurchenko2004,morishita2008,xu2012,wolter2016} and have been extended to condensed-matter phase in recent years \cite{vampa2014,vampa2015semiclassical,osika2017,yue2020,parks2020}.

\begin{figure}[!t]
	\includegraphics[width=12cm]{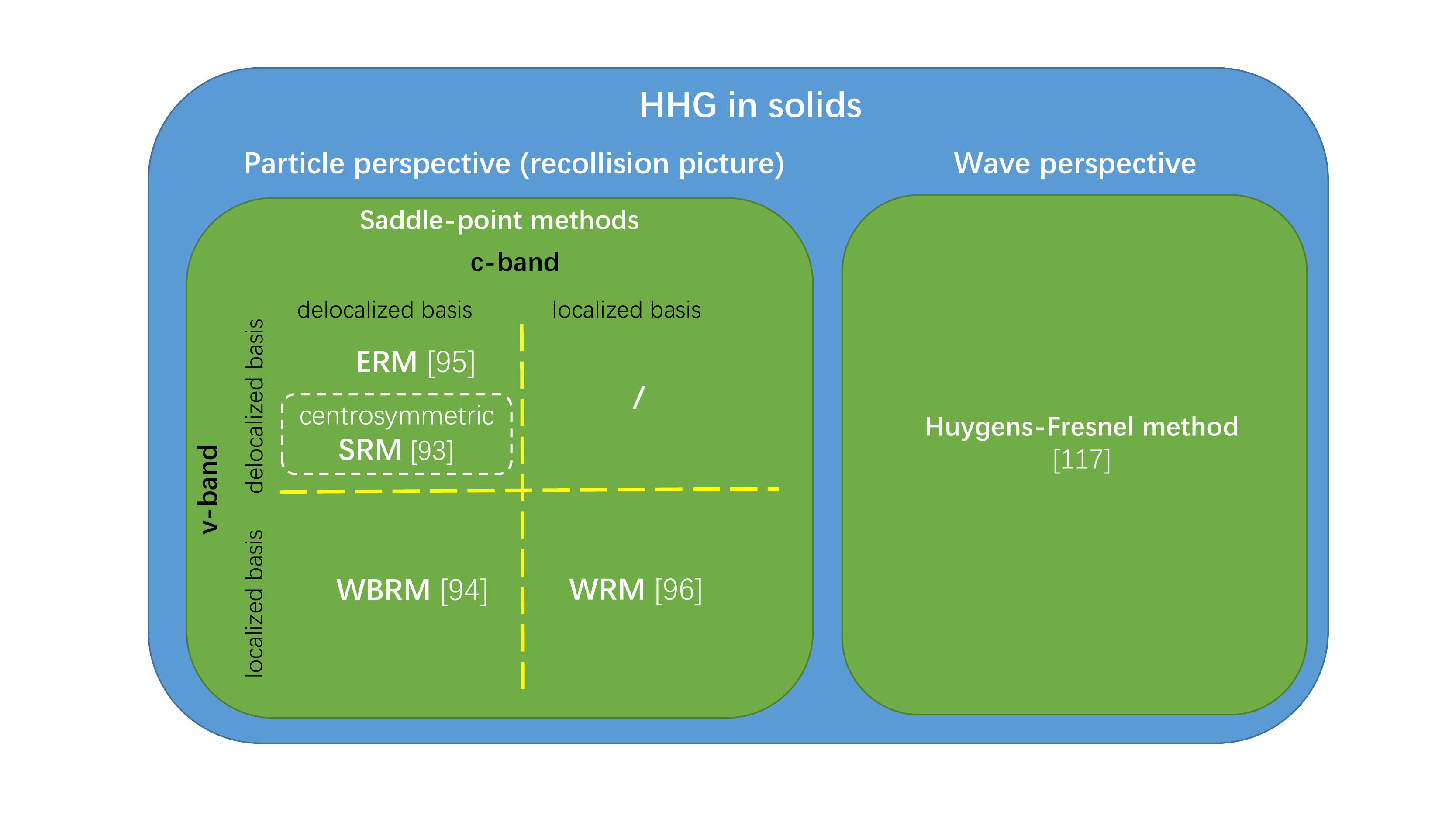}
	\caption{A summary of various types of methods for HHG in solids. ERM: extended recollision model, SRM: simple recollision model, WRM: Wannier recollision model, WBRM: Wannier-Bloch recollision model.}\label{methods}
\end{figure}

The first step to solving the TDSE is to choose a gauge and basis. The exact solution does not depend on this choice, but the chosen gauge and basis dictate approximations that one may wish to make, and they influence the physical interpretation of results. Here, we briefly summarize several typical analytical forms of interband current and the corresponding recollision models for HHG in solids based on saddle point methods (see Fig. \ref{methods}). Note that, the integral form for the currents, the derivation of the SBEs and the corresponding physical pictures are similar for different specific basis, e.g., employing the Bloch states or the Houston states. Thus, we will only in detail introduce some typical models classified based on the choice of delocalized or localized basis for the conduction and valence bands respectively, and focus on the discussions of saddle point equations and the corresponding recollision models. \\

\noindent \textbf{$\bullet$ Extended recollision model (ERM)}

We first continue with Eq. (35), which is derived from velocity gauge SBEs with Houston basis. By transforming into the frame $\textbf{k} = \textbf{k}_{0}+\textbf{A}(t)$, one can obtain the interband current in frequency domain as
\begin{align}\label{interband current for imperfect recollision}
J_{\mu}^{\mathrm{ERM}}(\omega)=\int_{-\infty}^{\infty} d t e^{i \omega t} \mathrm{J}_{\mu}^{\mathrm{ERM}}(t) = \sum_{\mathbf{k}} R_{\mu}^{\mathbf{k}} \int^{t}_{t_{0}} T^{\kappa(\textbf{k},t, s)} e^{-i\left[S^{\mu}(\mathbf{k}, t, s)-\omega t\right]} \mathrm{d} s+\text{c.c.}.
\end{align}
Here $\mu=\{x, y, z\}$ is the Cartesian indices, $T^{\kappa(\textbf{k},t, s)}=\text{F}(s)\left|\mathbf{d}_{\|}^{\kappa(\textbf{k},t, s)}\right|$ is the Rabi frequency,
$R_{\mu}^{\mathbf{k}}=\omega_{g}^{\mathbf{k}}\left|d_{\mu}^{\mathbf{k}}\right|$ is the recombination dipole, $\kappa\left(\textbf{k},t, t^{\prime}\right)=\mathbf{k}-\mathbf{A}(t)+\mathbf{A}\left(t^{\prime}\right)$ is the time-dependent crystal momentum.
\begin{align}\label{semiclassical action}
S^{\mu}(\mathbf{k}, t, s)=\int_{s}^{t}\left[\omega_{g}^{\kappa\left(\textbf{k},t, t^{\prime}\right)}+\mathbf{F}\left(t^{\prime}\right)\cdot\mathbf{\Lambda}_{cv}^{\kappa\left(\textbf{k},t, t^{\prime}\right)}\right] \mathrm{d} t^{\prime}+\beta_{\mu}^{\mathbf{k}}-\beta_{\|}^{\kappa(\textbf{k},t, s)}.
\end{align}
is the accumulated phase with
$\mathbf{\Lambda}_{cv}^{\textbf{k}}=\mathbf{\Lambda}_{c}^{\mathbf{k}}-\mathbf{\Lambda}_{v}^{\mathbf{k}}$ the Berry connection difference and $\beta_{\mu}^{\mathbf{k}} \equiv \arg \left(d_{\mu}^{\mathbf{k}}\right)$ the transition-dipole phases. The phase of the exponential $S^{\mu}(\mathbf{k}, t, s)-\omega t$ can vary wildly, introducing extreme cancellations in the integrand that require increased precision in the numerical integration to calculate accurately.

This problem can be overcome by employing the method of steepest descents for any relevant oscillatory integrals, where only the phase stationary (saddle) point contributes to the integrand while highly oscillatory terms are neglected. Thus,  the integrals can be approximated using the values of the integrand at stationary points of the action, reminiscent of the emergence of classical trajectories as the stationary-action points of the Feynman path integral.

For Eq. (\ref{interband current for imperfect recollision}), the major contribution to the integral over $\textbf{k}$ come from the stationary points determined by taking the partial derivatives with respect to $\textbf{k}$
\begin{align}\label{saddle point equation for k}
&\nabla_{\textbf{k}}\left[S^{\mu}(\mathbf{k}, t, s)-\omega t\right]   \nonumber\\
=&\nabla_{\textbf{k}}\left\{\int_{s}^{t}\left[\omega_{g}^{\kappa\left(\textbf{k},t, t^{\prime}\right)}+\mathbf{F}\left(t^{\prime}\right) \cdot \mathbf{\Lambda}_{cv}^{\kappa\left(\textbf{k},t, t^{\prime}\right)}\right] \mathrm{d} t^{\prime}+\beta_{\mu}^{\mathbf{k}}-\beta_{\|}^{\kappa(\textbf{k},t, s)}\right\}   \nonumber\\
=&\int_{s}^{t}\nabla_{\textbf{k}}\omega_{g}^{\kappa\left(\textbf{k},t, t^{\prime}\right)}\mathrm{d} t^{\prime}+\int_{s}^{t}\nabla_{\textbf{k}}\left[\mathbf{F}\left(t^{\prime}\right)\cdot\mathbf{\Lambda}_{cv}^{\kappa\left(\textbf{k},t, t^{\prime}\right)}\right] \mathrm{d} t^{\prime}+\nabla_{\textbf{k}}\beta_{\mu}^{\mathbf{k}}-\nabla_{\textbf{k}}\beta_{\|}^{\kappa(\textbf{k},t, s)}.
\end{align}
Using the relation $\nabla(\textbf{f}\cdot\textbf{g}) = \textbf{f}\times(\nabla\times\textbf{g})+(\textbf{f}\cdot\nabla)\textbf{g}+\textbf{g}\times(\nabla\times\textbf{f})+(\textbf{g}\cdot\nabla)\textbf{f}$ in the second term, one gets
\begin{align}\label{saddle point equation for k-1}
&\int_{s}^{t}\nabla_{\textbf{k}}\left[\mathbf{F}\left(t^{\prime}\right)\cdot\mathbf{\Lambda}_{cv}^{\kappa\left(\textbf{k},t, t^{\prime}\right)}\right] \mathrm{d} t^{\prime}   \nonumber\\
=& \int_{s}^{t}\textbf{F}(t')\times\left(\nabla_{\textbf{k}}\times\mathbf{\Lambda}_{cv}^{\kappa\left(\textbf{k},t, t^{\prime}\right)}\right)\mathrm{d} t^{\prime}
+\int_{s}^{t}\left(\textbf{F}(t')\cdot\nabla_{\textbf{k}}\right)\mathbf{\Lambda}_{cv}^{\kappa\left(\textbf{k},t, t^{\prime}\right)}\mathrm{d} t^{\prime}   \nonumber\\
=& \int_{s}^{t}\textbf{F}(t')\times\left(\mathbf{\Omega}_{c}^{\kappa(\textbf{k},t,t')}-\mathbf{\Omega}_{v}^{\kappa(\textbf{k},t,t')}\right)\mathrm{d} t^{\prime}-\mathbf{\Lambda}_{cv}^{\kappa(\textbf{k},t,t')}\bigg|_{t' = s}^{t'=t},
\end{align}
with the Berry curvature $\mathbf{\Omega}_{m}^{\textbf{k}} = \nabla_{\textbf{k}}\times\mathbf{\Lambda}_{m}^{\textbf{k}}$. Substituting Eq. (\ref{saddle point equation for k-1}) into (\ref{saddle point equation for k}) and reorganizing the formula, the saddle point condition with respect to the integration variable $\textbf{k}$ is
\begin{align}\label{saddle point equation for k-2}
\nabla_{\textbf{k}}\left[S^{\mu}(\mathbf{k}, t, s)-\omega t\right] = \Delta\textbf{r}(\textbf{k},t,s)-\Delta\textbf{D}(\textbf{k},t,s) = \textbf{0},
\end{align}
with the electron-hole separation vector and group velocities
\begin{align}\label{electron-hole separation vector}
\Delta\textbf{r}(\textbf{k},t,s) =& \int_{s}^{t}\left[\textbf{v}_{c}^{\kappa(\textbf{k},t,t')}-\textbf{v}_{v}^{\kappa(\textbf{k},t,t')}\right]\mathrm{d}t',   \\
\textbf{v}_{m}^{\kappa(\textbf{k},t,t')} =& \nabla_{\textbf{k}}E_{m}\textbf{(}\kappa(\textbf{k},t,t')\textbf{)}+\textbf{F}(t')\times\mathbf{\Omega}_{m}^{\kappa(\textbf{k},t,t')},
\end{align}
and the structure-gauge invariant displacement
\begin{align}\label{structure-gauge invariant displacement}
	\Delta\textbf{D}(\textbf{k},t,s) =& \textbf{D}_{\mu}^{\textbf{k}}-\textbf{D}_{\|}^{\kappa(\textbf{k},t,s)}   \\
	\textbf{D}_{\mu}^{\textbf{k}} =& \mathbf{\Lambda}_{cv}^{\textbf{k}}-\nabla_{\textbf{k}}\beta_{\mu}^{\textbf{k}}.
\end{align}
Therefore, one can conclude that the saddle points correspond to those crystal momenta $\textbf{k}_{\text{st}}$ for which the electron and hole are born at time $s$ and recombine at time $t$ with the electron-hole separation $\Delta\textbf{r}$ is equal to $\Delta\textbf{D}$. The physical meaning of the mathematical result expressed by Eq. (\ref{saddle point equation for k-2}) is clear: the dominant contribution of the harmonic emission comes from the electron-hole pairs tunneling away from but then recolliding with each other while oscillating in the laser field. In gas cases, it is evident that the electron-nucleus separation must be close to zero, because it is the only position where the transitions to and from the ground state can possibly occur, considering that the ground states are highly localized. However, the case in solids is distinct due to the nonzero $\Delta D$, reflecting the delocalized feature of the electrons in solids.

Following the same procedure, the major contribution to the integral over $s$ and $t$ come from the stationary points determined by taking the partial derivatives with respect to $s$ and $t$,
\begin{align}\label{saddle point equation for t and s}
\ &\partial_{s}\left[S^{\mu}(\mathbf{k}, t, s)-\omega t\right] = 0   \nonumber\\
\Leftrightarrow\ & \partial_{s}\left\{\int_{s}^{t}\left[\omega_{g}^{\kappa\left(\textbf{k},t, t^{\prime}\right)}+\mathbf{F}\left(t^{\prime}\right)\cdot\mathbf{\Lambda}_{cv}^{\kappa\left(\textbf{k},t, t^{\prime}\right)}\right] \mathrm{d} t^{\prime}-\beta_{\|}^{\kappa(\textbf{k},t, s)}\right\} = 0   \nonumber\\
\Leftrightarrow\ & -\omega_{g}^{\kappa\left(\textbf{k},t, s\right)}-\mathbf{F}\left(s\right)\cdot\mathbf{\Lambda}_{cv}^{\kappa\left(\textbf{k},t, s\right)}-\nabla_{\textbf{k}}\beta_{\|}^{\kappa(\textbf{k},t, s)}\partial_{s}\kappa(\textbf{k},t,s) = 0   \nonumber\\
\Leftrightarrow\ & \left[\omega_{g}^{\kappa\left(\textbf{k},t, s\right)}+\mathbf{F}(s)\cdot\textbf{D}_{\|}^{\kappa(\textbf{k},t,s)}\right] = 0,   \\
\nonumber\\
\ &\partial_{t}\left[S^{\mu}(\mathbf{k}, t, s)-\omega t\right] = 0   \nonumber\\
\Leftrightarrow\ &\partial_{t}\left\{\int_{s}^{t}\left[\omega_{g}^{\kappa\left(\textbf{k},t, t^{\prime}\right)}+\mathbf{F}\left(t^{\prime}\right)\cdot\mathbf{\Lambda}_{cv}^{\kappa\left(\textbf{k},t, t^{\prime}\right)}\right] \mathrm{d} t^{\prime}-\beta_{\|}^{\kappa(\textbf{k},t, s)}\right\}-\omega=0   \nonumber\\
\Leftrightarrow\  &\omega_{g}^{\textbf{k}}+\mathbf{F}\left(t\right)\cdot\mathbf{\Lambda}_{cv}^{\textbf{k}}+\int_{s}^{t}\partial_{t}\left[\omega_{g}^{\kappa\left(\textbf{k},t, t^{\prime}\right)}+\mathbf{F}\left(t^{\prime}\right)\cdot\mathbf{\Lambda}_{cv}^{\kappa\left(\textbf{k},t, t^{\prime}\right)}\right] \mathrm{d} t^{\prime}-\textbf{F}(t)\cdot\nabla_{\textbf{k}}\beta_{\|}^{\kappa(\textbf{k},t, s)}-\omega=0   \nonumber\\
\Leftrightarrow\  &\omega_{g}^{\textbf{k}}+\mathbf{F}\left(t\right)\cdot\left\{\mathbf{\Lambda}_{cv}^{\textbf{k}}-\nabla_{\textbf{k}}\beta_{\|}^{\kappa(\textbf{k},t, s)}+\int_{s}^{t}\nabla_{\textbf{k}}\left[\omega_{g}^{\kappa\left(\textbf{k},t, t^{\prime}\right)}+\mathbf{F}\left(t^{\prime}\right)\cdot\mathbf{\Lambda}_{cv}^{\kappa\left(\textbf{k},t, t^{\prime}\right)}\right] \mathrm{d} t^{\prime}\right\}-\omega=0   \nonumber\\
\Leftrightarrow\ &\omega_{g}^{\textbf{k}}+\mathbf{F}\left(t\right)\cdot\left\{\mathbf{\Lambda}_{cv}^{\textbf{k}}-\nabla_{\textbf{k}}\beta_{\|}^{\kappa(\textbf{k},t, s)}+\Delta\textbf{r}(\textbf{k},t,s)-\mathbf{\Lambda}_{cv}^{\kappa(\textbf{k},t,t')}\bigg|_{t' = s}^{t'=t}\right\}-\omega=0   \nonumber\\
\Leftrightarrow\ & \omega_{g}^{\textbf{k}}+\mathbf{F}\left(t\right)\cdot\left[\Delta\textbf{r}(\textbf{k},t,s)+\textbf{D}_{\|}^{\kappa(\textbf{k},t,s)}\right] = \omega. \label{saddle point equation for t}
\end{align}
In the fifth line of Eq. (46), we use the results in Eq. (\ref{saddle point equation for k-1}). The saddle points $(\textbf{k}_{\text{sp}},t_{\text{sp}},s_{\text{sp}})$ can be obtained by solving the Eqs. (\ref{saddle point equation for k-2}), (45), and (46), which are termed the tunneling, recollision, and emission equations, respectively. Then, within the saddle point approximation, the interband currents are given by a sum over all the relevant saddle point roots. Following the saddle point equations, the interband HHG can be interpreted in terms of the following three steps: i) the tunneling Eq. (45) gives the tunneling conditions $\textbf{k}' = \textbf{k}_{\text{sp}}+\textbf{A}(t_{\text{sp}})-\textbf{A}(s_{\text{sp}})$, that is, an electron-hole pair is created by tunneling from VB to CB at time $s_{\text{sp}}$ and with an initial momentum $\textbf{k}'$; ii) the electron and hole are accelerated by the laser and the recollision equation Eq. (\ref{saddle point equation for k-2}) constrains the relation between ionization time $t_{\text{sp}}$ and emission time $s_{\text{sp}}$, i.e., the instant the electron-hole separation $\Delta\textbf{r}(\textbf{k}_{\text{sp}},t_{\text{sp}},s_{\text{sp}})$ is equal to $\Delta\textbf{D}(\textbf{k}_{\text{sp}},t_{\text{sp}},s_{\text{sp}})$; iii) the emission frequency $\omega$ due to the electron-hole pair recombination at time $t_{\text{sp}}$ with crystal momentum $\textbf{k}_{\text{sp}}$ and relative separation $\Delta\textbf{r}$ is determined by Eq. (\ref{saddle point equation for t}), which is simply the energy conservation law. This three-step model is referred to as an extended recollision model (ERM) by relaxing the recollision condition by a preset recollision threshold $R_{0}$, which is introduced by Yue and Gaarde's papers \cite{yue2020,yue2022}. In ERM, a semiclassical recollision event is recorded when $\Delta\text{R}^{\mu} = \left|\Delta\textbf{r}(\textbf{k},t,s)-\Delta\textbf{D}(\textbf{k},t,s)\right|$ reaches its minimum within the region $\Delta\text{R}^{\mu}<\text{R}_{0}$. \\

\noindent \textbf{$\bullet$ Simple recollision model (SRM)}

For a direct band-gap centrosymmetric material, the interband current in Eq. (\ref{interband current for imperfect recollision}) is deduced as
\begin{align}\label{interband current for SRM}
J_{\mu}^{\mathrm{SRM}}(\omega)=\omega\sum_{\mathbf{k}} \left[\textbf{d}_{\mu}^{\textbf{k}}\right]^{*}\int_{-\infty}^{+\infty}\mathrm{d}t e^{i\omega t}\int_{t_{0}}^{t}\mathrm{d}s\textbf{F}(s)\cdot\textbf{d}^{\kappa(k,t,s)}e^{-i\int_{s}^{t}\omega_{g}^{\kappa(\textbf{k},t,t')}\mathrm{d}t'},
\end{align}
and the corresponding saddle point equations have a simplified form
\begin{align}\label{saddle points for SRM}
&\Delta\textbf{r}(\textbf{k},t,s) = \int_{s}^{t}\nabla_{\textbf{k}}\omega_{g}^{\kappa(\textbf{k},t,t')}dt' = 0,   \\
&\omega_{g}^{\kappa(\textbf{k},t,s)} = 0,   \\
&\omega_{g}^{\textbf{k}}+\textbf{F}(t)\cdot\Delta\textbf{r}(\textbf{k},t,s) = \omega.
\end{align}
These results are also proposed by Vampa et al. using the length gauge SBEs with a Bloch basis. Note that the term $\textbf{F}(t)\cdot\Delta\textbf{r}(\textbf{k},t,s)$ in Eq. (50) is omitted in Vampa's work \cite{vampa2015semiclassical} considering the recollision condition Eq. (48). Neglecting the imaginary parts of the saddle point, the saddle point conditions give the simple recollision model (SRM) similar to the three-step model of atomic HHG \cite{corkum1993,vampa2015semiclassical} : (i) an electron-hole pair is created with zero crystal momentum at time $s$ (at the $\Gamma$ point); (ii) the electron and hole are separated by the laser field with the instantaneous velocity $\nabla_{\textbf{k}}\omega_{g}^{A(t)-A(t')}$; (iii) the electron and hole re-encounter to each other at time $t$ and release a harmonic photon with energy $\omega = \omega_{g}^{\textbf{k}}$. \\

\noindent \textbf{$\bullet$ Wannier recollision model (WRM)}

By transforming the interband current [Eq. (\ref{interband current for SRM})] from Bloch to Wannier basis followed by saddle-point integration, Parks et al. develop a generalized quasi-classical approach accounting for the lattice structure \cite{parks2020}. The connection between the Bloch and Wannier basis functions is given by a Fourier transform according to
\begin{align}\label{Wannier functions}
u_{m, \mathbf{k}}(\mathbf{x})=\sum_{j} w_{m}\left(\mathbf{x}-\mathbf{x}_{j}\right) e^{-i \mathbf{k} \cdot\left(\mathbf{x}-\mathbf{x}_{j}\right)},   \\
w_{m}\left(\mathbf{x}-\mathbf{x}_{j}\right)=\frac{1}{V_{\text{cell}}} \int_{\mathrm{BZ}} u_{m, \mathbf{k}}(\mathbf{x}) e^{i \mathbf{k} \cdot\left(\mathbf{x}-\mathbf{x}_{j}\right)} \mathrm{d} \mathbf{k}.
\end{align}
Here, $w_{m}\left(\mathbf{x}-\mathbf{x}_{j}\right)$ is the Wannier function of band $m$ corresponding to the primitive unit cell at position $\textbf{x}_{j}$. $V_{\text{cell}}$ is the volume of a unit cell, and the integration is performed over the first Brillouin zone (BZ). As all lattice sites are identical, it is sufficient to investigate $\textbf{x}_{j} = 0$, and the initial wave function can be written as
\begin{align}\label{initial wave function}
\Psi(\mathbf{x}, 0)=\int_{\mathrm{BZ}} \mathrm{d} \mathbf{k} \psi_{v, \mathbf{k}}(\mathbf{x}) a_{v}(\mathbf{k}, t=0)=w_{v}(\mathbf{x}).
\end{align}
The lattice structure is involved by transforming the Bloch dipole moment $\textbf{d}(\textbf{k})$ to Wannier dipole moment $\textbf{d}_{l} = \int_{V_{\text{crystal}}} w_{c}^{*}\left(\mathbf{x}-\mathbf{x}_{l}\right) \mathbf{x} w_{v}(\mathbf{x}) \mathrm{d} \mathbf{x}$ with Wannier functions
\begin{align}\label{TDM in Wannier function}
\mathbf{d}(\mathbf{k}) &=\sum_{j, k} \int_{V_{\text{cell}}} w_{c}^{*}\left(\mathbf{x}-\mathbf{x}_{k}\right)\left[\mathbf{x}-\mathbf{x}_{j}\right] w_{v}\left(\mathbf{x}-\mathbf{x}_{j}\right) e^{i \mathbf{k} \cdot\left(\mathbf{x}_{j}-\mathbf{x}_{k}\right)} \mathrm{d} \mathbf{x}   \nonumber\\
&=\sum_{j, l} \int_{V_{\text{cell}}} w_{c}^{*}\left(\mathbf{x}-\left(\mathbf{x}_{j}+\mathbf{x}_{l}\right)\right)\left[\mathbf{x}-\mathbf{x}_{j}\right] w_{v}\left(\mathbf{x}-\mathbf{x}_{j}\right) e^{-i \mathbf{k} \cdot \mathbf{x}_{l}} \mathrm{~d} \mathbf{x}   \nonumber\\
&=\sum_{l} e^{-i \mathbf{k} \cdot \mathbf{x}_{l}} \int_{V_{\text{crystal}}} w_{c}^{*}\left(\mathbf{x}-\mathbf{x}_{l}\right) \mathbf{x} w_{v}(\mathbf{x}) \mathrm{d} \mathbf{x}=\sum_{l} \mathbf{d}_{l} e^{-i \mathbf{k} \cdot \mathbf{x}_{l}}.
\end{align}
In the above derivation, a transform $\mathbf{x}_{k}=\mathbf{x}_{j}+\mathbf{x}_{l}$ changes the summation index $k$ with $l$ in the second line, and performing $\sum_{j}$ changes the integration volume from a unit cell to the whole crystal in the second line. The Wannier dipole moment $\mathbf{d}_{l}$ describes a transition where an electron is born $l$ lattice away from the hole. Inserting Eq. (\ref{TDM in Wannier function}) into Eq. (\ref{interband current for SRM}), one can obtain the real-space interband current,
\begin{align}\label{Interband current for WRM}
J_{\mu}^{\mathrm{WRM}}(\omega) &=\sum_{j, l}\left\{\mathbf{d}_{j}^{*}\left[\mathbf{d}_{l} \cdot \mathbf{T}_{j l}(\omega)\right]-\mathbf{d}_{j}\left[\mathbf{d}_{l}^{*} \cdot \mathbf{T}_{j l}^{*}(-\omega)\right]\right\} \nonumber\\
&=\sum_{j, l}\left[\mathbf{P}_{j l}(\omega)-\mathbf{P}_{j l}^{*}(-\omega)\right],   \\
\mathbf{T}_{j l}(\omega) &=\omega\sum_{\mathbf{k}}\int_{-\infty}^{\infty} \mathrm{d} t \int_{-\infty}^{t} \mathrm{d}s \mathbf{F}\left(s\right) e^{i \varphi\left(\mathbf{k},t,s,\mathbf{x}_{j},\mathbf{x}_{l}\right)}.
\end{align}
where $\varphi=-\int_{s}^{t}\omega_{g}^{\kappa(k,t,t')}\mathrm{d}t'+\omega t+\mathbf{k} \cdot\left(\mathbf{x}_{j}-\mathbf{x}_{l}\right)+\left[\mathbf{A}(t)-\mathbf{A}\left(s\right)\right] \cdot \mathbf{x}_{l}$. These integrals are solved by saddle-point integration with saddle point conditions
\begin{align}\label{saddle point for WRM}
\Delta\textbf{r}(\textbf{k},t,s) = \textbf{x}_{j}-\textbf{x}_{l},   \\
\omega_{g}^{\kappa(\textbf{k},t,s)}+\textbf{F}(s)\cdot\textbf{x}_{l} = 0,   \\
\omega_{g}^{\textbf{k}}+\textbf{F}(t)\cdot\textbf{x}_{j} = \omega,
\end{align}
resulting from the partial derivatives with $\varphi$ respect to the partial derivatives with respect to $\textbf{k}$, $s$, and $t$. This Wannier quasi-classical description transforms the three-step model from reciprocal space to real space \cite{parks2020}: (i) an electron is transitioned from $\textbf{x}_0$ to $\textbf{x}_0 + \textbf{x}_l$ at birth time $s$; (ii) the electron-hole pair is separated by the laser field following the classical trajectory $\Delta\textbf{r}(\textbf{k},t,s)$; (iii) the electron and hole recombine at time $t$ with probability amplitude $\mathbf{d}_{j} e^{-i \mathbf{k}_{s}\left(t_{b}, t_{r}\right) \cdot \mathbf{x}_{j}}$ and release a harmonic photon of frequency $\omega = \omega_{g}^{\textbf{k}}+\textbf{F}(t)\cdot\textbf{x}_{j}$ when the separation $\Delta\textbf{r}(\textbf{k},t,s)$ is equal to $\textbf{x}_{j}-\textbf{x}_{l}$. \\

\noindent \textbf{$\bullet$ Wannier-Bloch recollision model}

The Wannier wave functions can also be used as a basis for the numerical analysis of HHG. A Wannier-Bloch approach was first developed by Osika et al. in their article \cite{osika2017}, which assesses the contributions of individual lattice sites to the HHG process and hence precisely addresses the question of localization of harmonic emission in solids. In the Wannier-Bloch approach, the TDSE is solved with the ansatz
\begin{align}\label{Wave function for Wannier-Bloch approach}
|\Psi(t)\rangle=\sum_{j}\left|w_{v, j}\right\rangle a_{j}(t)+\int_{\mathrm{BZ}} a_{c}(\textbf{k}, t)\left|\psi_{c, k}\right\rangle\mathrm{d}\textbf{k},
\end{align}
where the wave function of a single electron is expressed as a superposition of the localized Wannier wave states $\left|w_{v, j}\right\rangle$ in the valence band and delocalized Bloch state $\left|\psi_{c, k}\right\rangle$ in the conduction band. Here, $j$ runs over all atomic sites in the crystal. The Wannier functions of an $m$th band can be represented by a set of Bloch functions
\begin{align}\label{Wannier basis}
\left|w_{m, j}\right\rangle=w_{m, j}(x)=\int_{\mathrm{BZ}}\psi_{m, \textbf{k}}\left(x-x_{j}\right) \tilde{w}_{m}(\textbf{k})\mathrm{d}\textbf{k},
\end{align}
In this form, $\tilde{w}_{m}(\textbf{k})$ is a product of a normalization constant and a phase depending on crystal momentum $\textbf{k}$. The discussion is restricted to Wannier functions under appropriate symmetry and falling off exponentially with distance. In this case, the $\tilde{w}_{m}(\textbf{k})$ are set to be independent of $k$. Note that the Wannier functions form a completely orthogonal set in the valence band but are not eigenfunctions of the Hamiltonian, the electron wave function may spread in the lattice because of the interatomic hopping and the acceleration driven by the laser field. Thus, the coefficient $a_{j}(t)$ acquires nonzero values during the laser pulse, and the harmonic emission is obtained by summing up the contribution of each Wannier state
\begin{align}\label{Interband current for WBRM}
J_{\mu}^{\mathrm{WBRM}}(\omega) &= \omega\int_{-\infty}^{+\infty}\mathrm{d}te^{i\omega t}\sum_{\textbf{k}}\sum_{j}a_{j}^{*}(t)\textbf{d}_{jc}(\textbf{k})a_{c}(\textbf{k},t)+c.c.,   \nonumber\\
&= \omega\left|\tilde{w}_{v}\right|^{2}\int_{-\infty}^{+\infty}\mathrm{d}te^{i\omega t}\sum_{\mathbf{k}}\sum_{j}a_{j}^{*}(t)\left[\textbf{d}_{\mu}^{\textbf{k}}\right]^{*}e^{i\textbf{k}\cdot\textbf{x}_{j}}   \nonumber\\
& \ \ \ \times\sum_{j'}\int_{t_{0}}^{t}\mathrm{d}s\textbf{F}(s)\cdot\textbf{d}^{\kappa(\textbf{k},t,s)}e^{-i\kappa(\textbf{k},t,s)\cdot\textbf{x}_{j'}}e^{-i\int_{s}^{t}E_{c}\textbf{(}\kappa(\textbf{k},t,t')\textbf{)}\mathrm{d}t'}.
\end{align}
The transition dipole moment from the conduction band to the valence band is $\textbf{d}_{jc}(\textbf{k}) = \textbf{d}_{vc}(\textbf{k})\tilde{w}_{v}^{*}e^{i\textbf{k}\cdot\textbf{x}_{j}}$. Then, one can reorganize this current as a product of a global amplitude and phase
\begin{align}\label{Interband current for WBRM-1}
J_{\mu}^{\mathrm{WBRM}}(\omega) &= \sum_{j,j'}\sum_{\mathbf{k}}\int_{-\infty}^{+\infty}\mathrm{d}t\int_{t_{0}}^{t}\mathrm{d}sf_{j,j'}(\textbf{k},t,s)e^{-i\Phi_{j,j'}(\textbf{k},t,s)+i\omega t},   \\
f_{j,j'}(\textbf{k},t,s) &= \omega\left|\tilde{w}_{v}\right|^{2}\left|a_{j}(t)\right|\left[\textbf{d}_{\mu}^{k}\right]^{*}\left|a_{j'}(s)\right|\textbf{F}(s)\cdot\textbf{d}^{\kappa(\textbf{k},t,s)},   \\
\Phi_{j,j'}(\textbf{k},t,s) &= \int_{s}^{t}E_{c}\textbf{(}\kappa(\textbf{k},t,t')\textbf{)}\mathrm{d}t'+\varphi_{a_{j}}(t)-\textbf{k}\cdot\textbf{x}_{j}-\varphi_{a_{j'}}(s)+\kappa(\textbf{k},t,s)\cdot\textbf{x}_{j'}
\end{align}
The physical information about the radiation by means of the Wannier-Bloch approach is extracted from the saddle point conditions
\begin{align}\label{saddle point equation for WBRM-2}
\int_{s}^{t}\nabla_{\textbf{k}}E_{c}\textbf{(}\kappa(\textbf{k},t,t')\textbf{)}\mathrm{d}t'+\textbf{x}_{j'}-\textbf{x}_{j} = 0,   \\
E_{c}\textbf{(}\kappa(\textbf{k},t,s)\textbf{)}+\partial_{s}\varphi_{a_{j'}}(s)+\textbf{F}(s)\cdot\textbf{x}_{j'} = 0,   \\
E_{c}(\textbf{k})+\partial_{t}\varphi_{a_{j}}(t)+\textbf{F}(t)\cdot\textbf{x}_{j} = \omega,
\end{align}
which gives a lattice site resolved recollision model \cite{osika2017}: (i) an electron located at the atomic site $\textbf{x}_{j'}$ is excited from the valence to the conduction band at time $s$; (ii) this electron is accelerated in the conduction band while the electron wave function spreads along the periodic lattice structure; (iii) at time $t$, the electron recombines at the site $\textbf{x}_{j}$, and the excess electron energy is emitted in the form of a high-harmonic photon.

\subsection{C. Discussions on saddle point approximation}

Since all the above-mentioned models root in the saddle point approximation, to discuss the scope of application of the model and the recollision picture, it is valuable to reexamine the fundamental of the validity of the saddle point approximation. Thus, we first recall a general form of saddle point integration theorem. We only quote some of the derivation processes and conclusions that will be called upon in this review. More complete and rigorous discussions about the saddle point method can be found in Refs. \cite{Nayak2019,Amini2019}. The saddle point approximation is a powerful method for analyzing the asymptotic behavior of a complex integration
\begin{align}\label{A general complex space integration}
I(\sigma) = \int_{C}f(z)e^{\sigma g(z)}\mathrm{d}z,
\end{align}
with parameter $\sigma\gg1$. Both $f(z)$ and $g(z)$ are smooth complex analytical functions of $z$ and $C$ is the integral path. We consider non-degenerate multiple isolated stationary phase points $z = z_{s}$ (index $s$) where $g'(z) = \nabla_{z}g(z)|_{z = z_{s}} = 0$ and $g''(z_{s})\neq0$. The integration path $C$ can be deformed to follow an appropriate path through the critical points of the integrand utilizing the Cauchy-Goursat theorem \cite{Moore1900} without changing the value of the integral. This allows one to make a very useful simplification in calculating the interested integral by expanding the exponential in the integrand along the steepest descent into a truncated Taylor series around the stationary phase point,
\begin{align}\label{Gaussian form integration}
I(\sigma) &= \int_{C}f(z)e^{\sigma g(z_{s})+\frac{1}{2}\sigma g''(z_{s})(z-z_{s})^{2}+...}\mathrm{d}z,   \nonumber\\
&\approx e^{\sigma g(z_{s})}\int_{C}f(z)e^{\frac{1}{2}\sigma g''(z_{s})(z-z_{s})^{2}}\mathrm{d}z.
\end{align}
Retaining to the second order, we get a Gaussian function with width $O(\sqrt{1/\sigma})$. Using the generalized Riemann-Lebesgue lemma, the contributions for this type of integral come predominantly from the stationary phase points $z_{s}$, while the oscillatory parts of the integrand cancel out for large $\sigma$. Then, the integration Eq. (\ref{A general complex space integration}) can be asymptotically approximated as a sum of stationary phase point contributions
\begin{align}\label{saddle point integration}
I(\sigma) &= \sum_{s}\left(\frac{2\pi}{\sigma}\right)^{\frac{1}{2}}\frac{f(z_{s})}{\sqrt{\det(-g''(z_{s}))}}e^{\sigma g(z_{s})}.
\end{align}
Equation (\ref{saddle point integration}) is called the saddle point approximation and the expression $\nabla_{z}g(z)|_{z = z_{s}} = 0$ is referred to as the saddle point equation (correspondingly, $z_{s}$ is called the saddle point).

\begin{figure}[!t]
	\includegraphics[width=17cm]{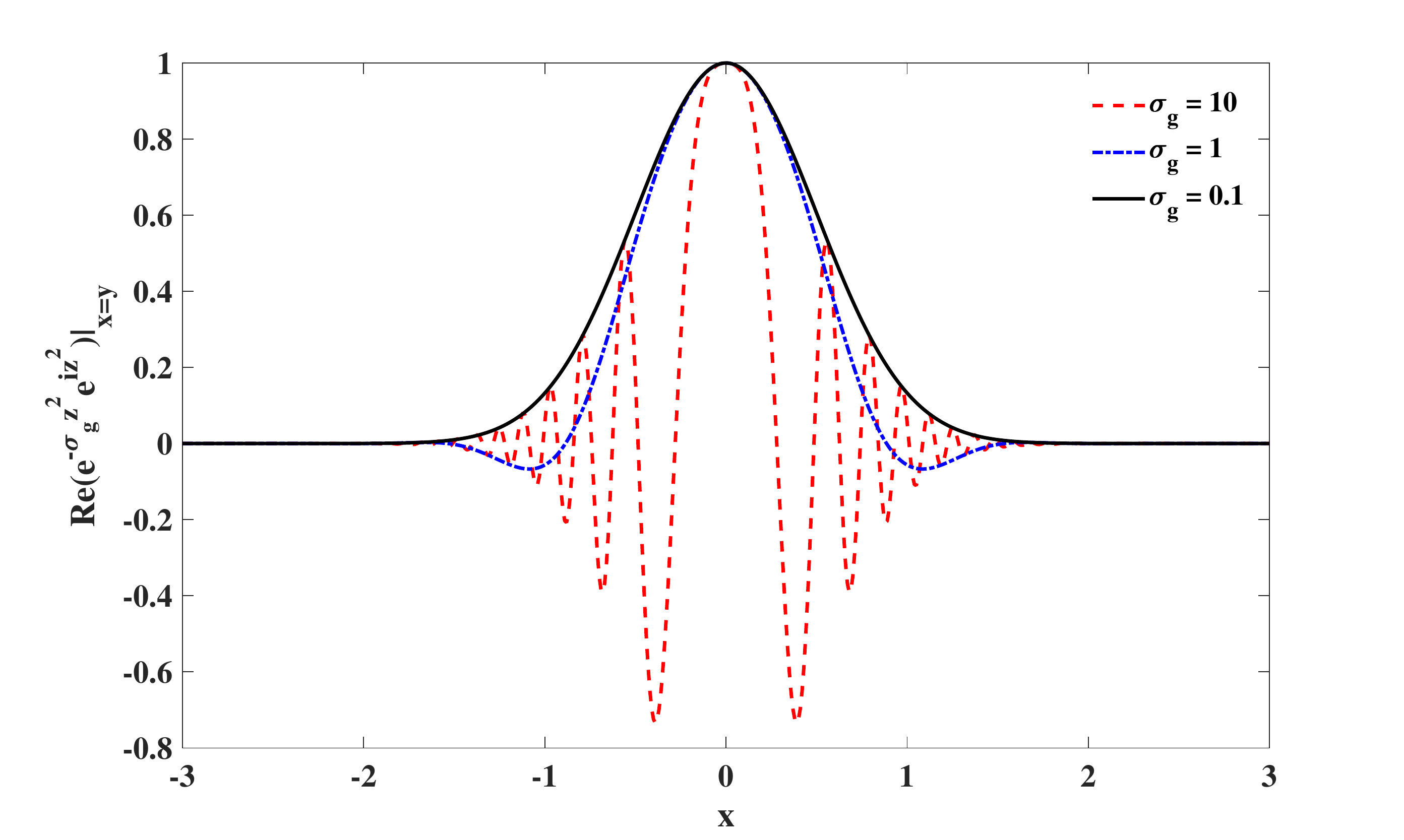}
	\caption{Behavior of the integrand of $I(\sigma_{g}) = \text{Re}\left(\int_{-\infty}^{+\infty}e^{-\sigma_{g}\sigma z^{2}}e^{i\sigma z^{2}}\mathrm{d}z\right)$ around the saddle point along the steepest descent with different parameter $\sigma_{g}$ and $\sigma=1$. The results with $\sigma_{g}<0.1$ have no discernible difference from the results with $\sigma_{g}=0.1$.}\label{saddle_point_behavior}
\end{figure}

From the above discussion, the saddle point approximation is valid only when $\sigma$ is large enough. To give an intuitive illustration, we take a typical form of integral $I(\sigma,\sigma_{g}) = \int_{-\infty}^{+\infty}e^{-\sigma_{g}\sigma x^{2}}e^{i\sigma x^{2}}\mathrm{d}x$ as an example, where $e^{-\sigma_{g}\sigma x^{2}}$ and $e^{i\sigma x^{2}}$ corresponds to amplitude term $f(z)$ and phase term $e^{\sigma g(z)}$ in Eq. (\ref{A general complex space integration}) respectively. Utilizing Cauchy-Goursat theorem \cite{Moore1900}, one can deform the integration path to follow the steepest descent, i.e., $z = (x+iy) |_{x=y}$. In this form, for a fixed $\sigma$, a smaller $\sigma_{g}$ describes a relatively slowly varying amplitude compared with the phase in the integrand around the saddle point $x = 0$. Figure \ref{saddle_point_behavior} shows the behavior of $e^{-\sigma_{g}\sigma x^{2}}e^{i\sigma x^{2}}$ along the steepest descent, where the parameter $\sigma$ is set to 1 without loss of generality. Comparing the results for three different values of $\sigma_{g}$, it is shown that with decreasing $\sigma_{g}$ the evaluation of the oscillatory integral converges to a Gaussian form contributed by the phase term, i.e., $I(\sigma,\sigma_{g}) = \int_{-\infty}^{+\infty}e^{-\sigma_{g}\sigma x^{2}}e^{i\sigma x^{2}}\mathrm{d}x \approx \int_{-\infty}^{+\infty}e^{i\sigma x^{2}}\mathrm{d}x = \sqrt{\frac{2\pi}{-i\sigma}}$. Namely, the saddle point approximation (Eq. (\ref{A general complex space integration})) works well. However, for cases the oscillation of the amplitude term is comparable to ($\sigma_{g} = 1$) or even faster than ($\sigma_{g} = 10$) the phase term, the integration deviates from the result of saddle point approximation. This conclusion can be also obtained with an analytical form $I(\sigma,\sigma_{g}) = \int_{-\infty}^{+\infty}e^{-(\sigma_{g}-i)\sigma x^{2}}\mathrm{d}x = \sqrt{\frac{2\pi}{(\sigma_{g}-i)\sigma}}\xrightarrow{\lim{\sigma_{g}\rightarrow0}}\sqrt{\frac{2\pi}{-i\sigma}}$. Regarding the physical problems we are interested in, the coefficient in the phase term $\sigma$ is related to the electron energy. Thus, one can expect that the saddle-point integral gives a reasonable approximation when the accumulated electron energy is large enough and the integrand phase term varies much faster than the other factors with respect to the integration variables.

Combing back to Eq. (\ref{interband current for imperfect recollision}), the HHG contributed by the interband current is expressed as an integral with amplitude and propagation phase terms. Within the saddle point approximation, the interband current in Eq. (\ref{interband current for imperfect recollision}) is given by a sum over all the relevant stationary values $\{\textbf{k}_{\text{sp}},t_{\text{sp}},s_{\text{sp}}\}$
\begin{align}\label{interband current with saddle point k_st}
J_{\mu}^{\mathrm{ter}}(\omega)\approx\sum_{\textbf{k}_{\text{sp}},t_{\text{sp}},s_{\text{sp}}}H(\textbf{k}_{\text{sp}},t_{\text{sp}},s_{\text{sp}})R_{\mu}^{\mathbf{k}_{\text{st}}}T^{\textbf{k}_{\text{sp}}-\textbf{A}(t_{\text{sp}})+\textbf{A}(s_{\text{sp}})} e^{-i\left(S^{\mu}(\textbf{k}_{\text{sp}},t_{\text{sp}},s_{\text{sp}})-\omega t_{\text{sp}}\right)}+\text{c.c.}
\end{align}
with an additional Hessian factor $H(\textbf{k}_{\text{sp}},t_{\text{sp}},s_{\text{sp}}) = \frac{1}{\sqrt{\det\left[2\pi i\partial_{2}S^{\mu}|_{\textbf{k}_{\text{sp}},t_{\text{sp}},s_{\text{sp}}}\right]}}$ that accounts for the width of the complex-integration Gaussians being approximated. The term $\partial_{2}S^{\mu}$ is the Hessian matrix with element $\partial_{l}\partial_{m}S^{\mu}$ ($l$, $m = \{\textbf{k},t,s\}$). As discussed in Section 2B, the semiclassical picture of HHG is obtained in the scenario that the integral is approximated using the values of the integrand at stationary points of the phase term, based on which the classic correspondence between the electron trajectory and harmonic emission is established. Thus, the validity of the semiclassical picture rests on whether the saddle point approximation is valid.

In atomic gases, the integral form of harmonic emission under strong field approximation involves a highly oscillatory term in the phase factor $\Theta(\textbf{p},t,s) = \omega t - \int_{s}^{t}\mathrm{d}\tau\left(\frac{\left[\textbf{p}+\textbf{A}(t)\right]^{2}}{2}+I_{p}\right)$ \cite{lewenstein1994} with the canonical momentum $\textbf{p}$ and atomic ionization potential $I_{p}$. One can examine the behavior of the phase in the simplest case where an intense monochromatic linearly polarized laser field $\textbf{F}(t) = \textbf{F}_{0}\cos(\omega_{0}t)$ is applied
\begin{align}\label{phase term in gases}
\Theta(\textbf{p},t,s) &= \omega t-\int_{\omega s}^{\omega_{t}}\left(\frac{\left[p_{\|}-\frac{F_{0}}{\omega_{0}}\sin(\theta)\right]^2}{2}+I_{p}\right)\frac{\mathrm{d}\theta}{\omega_{0}}   \nonumber\\
&= \omega t-\frac{F_{0}^{2}}{2\omega_{0}^{3}}\int_{\omega s}^{\omega_{t}}\left[\frac{\omega_{0}p_{\|}}{F_{0}}-\sin(\theta)\right]\mathrm{d}\theta+\frac{I_{p}}{\omega_{0}}\left(\omega_{0}t-\omega_{0}s\right)   \nonumber\\
&= \omega t-\frac{U_{p}}{\omega_{0}}\int_{\omega s}^{\omega_{t}}\left[\frac{\omega_{0}p_{\|}}{F_{0}}-\sin(\theta)\right]\mathrm{d}\theta+\frac{I_{p}}{\omega_{0}}\left(\omega_{0}t-\omega_{0}s\right).
\end{align}
$U_{p} = \frac{F_{0}^{2}}{4\omega_{0}^{2}}$ is the ponderomotive energy of the electron under the action of the laser field, and $\frac{I_{p}}{\omega_{0}}$ is the number of photons necessary to ionize the target atom. The saddle point approximation is asymptotically exact provided $\frac{U_{p}}{\omega_{0}}$ and $\frac{I_{p}}{\omega_{0}}$ are large enough, which is in general well satisfied with a strong low-frequency laser field.

The condition is however quite different for the interband HHG in solids. On the one hand, in contrast to the free-electron dispersion relevant for gas HHG, solid systems have more complicated electron band structures. Then, the variation of the integrand phase term is actually dependent on a coupling form of the laser and crystal parameters, and one needs to meticulously evaluate the applicability of the saddle point approximation. On the other hand, the delocalization of electrons in solids indicates that the electron dynamics may not be fully described by the particle picture, and wave-like properties should be involved. To shed light on these problems, we consider a representative example by considering the evolution of a wave packet
\begin{align}\label{Wave_packet}
\Psi(x,t,t_{0}) = \int_{-\infty}^{\infty}f(k)e^{ikx-i\varphi(k,t,t_{0})}\mathrm{d}k,
\end{align}
where $f(k) = e^{-(k/k_{w})^{2}}$ is the Gaussian wavepacket in $k$-space, $\varphi(k,t,t_{0}) = \int_{t_{0}}^{t}d\tau\varepsilon(k+A(\tau)-A(t_{0}))$ is the dynamical phase with energy dispersion $\varepsilon$, and $A(t) = -F_{0}/\omega_{0}\text{sin}(\omega_{0}t)$ is the vector potential of the laser field. The saddle point approximation gives a classical picture that the center of the wave packet moves with group velocity $\nabla_{k}\varepsilon$, i.e., $x_{c}(t) = \int_{t_{0}}^{t}d\tau\nabla_{k}\varepsilon(k+A(\tau)-A(t_{0}))$.

\begin{figure}[!t]
	\includegraphics[width=17cm]{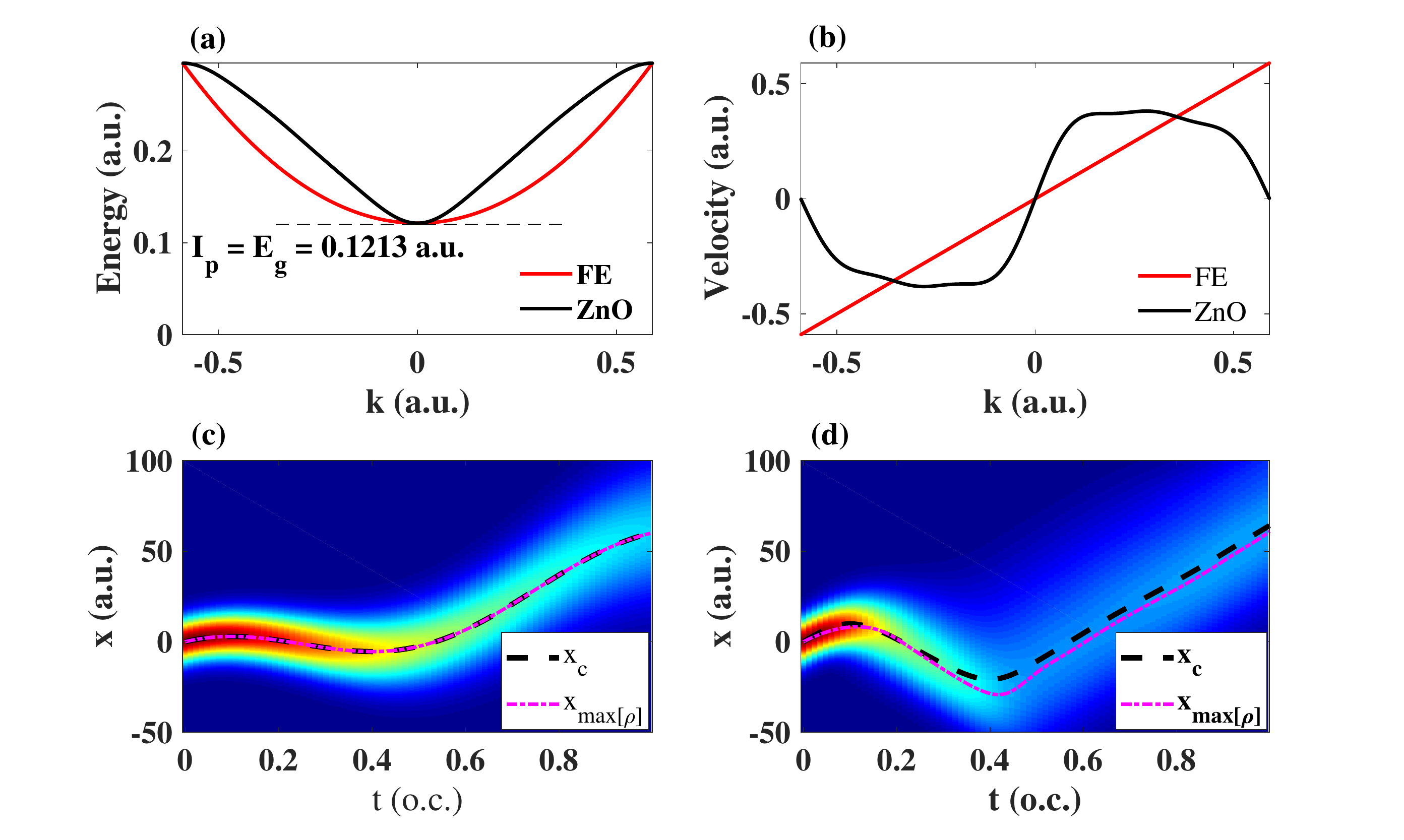}
	\caption{A numerical demonstration of the evolution of a wave packet with different energy dispersion. (a) The energy dispersions $\varepsilon(k)$ and (b) the corresponding velocity dispersions $\nabla_{k}\varepsilon(k)$ of FE and crystal ZnO. (c) and (d) in color show the time-dependent distributions of the wavepacket for FE and crystal ZnO, respectively. The maximum value of the wave function is marked by dashed pink lines, while the classical movement is marked by dashed black lines.}\label{wave_propagation}
\end{figure}

We evaluate the results from Eq. (\ref{Wave_packet}) with different dispersions: the free-electron (FE) dispersion (i.e., parabolic band) and a crystal energy dispersion (we take the first conduction band of ZnO crystal along the $\Gamma-M$ direction as an example). We use the same laser parameters as those in Refs. \cite{vampa2014,Li_2021}, and shift the parabolic band to mimic the same band gap of ZnO, i.e., $I_{p} = E_{g} = 0.1213$ a.u.. The energy and corresponding velocity dispersions are shown in Figs. \ref{wave_propagation} (a) and (b), respectively. One can see obvious nonlinear velocity dispersion for crystal ZnO in the region far away from the high-symmetry point. Using Eq. (\ref{Wave_packet}), one can obtain the time-dependent distribution of the wavepacket $\rho(x,t) = \Psi(x,t,t_{0})^{*}\Psi(x,t,t_{0})$. The results are shown in Figs. \ref{wave_propagation} (c) and (d) for FE and crystal ZnO. The conformal contour indicates the time-dependent distribution of the wavepacket and the dashed black lines indicate the classical movement $x_{c}$. In the case of FE, one can see that the shape of the wavepacket is almost unchanged except for a free diffusion. Besides, the classical movement is always consistent with the center of the wavepacket (the locations for maximum wavepacket distribution marked by dashed pink lines). However, for a crystal energy dispersion, one can clearly see that the Gaussian wavepacket is distorted during the evolution, and the classical movement fails to describe the locations of the maximum wavepacket distribution. Note that the discrepancy between $x_c$ and $x_{\max[\rho]}$ is closely related to the fact that the electron wavepacket distorts during evolution, which is typically a wave feature and cannot be described by treating an electron as a particle.

Before we proceed, let us provide a brief recap of previous discussions. Literatures have shown that the recollision models have been widely applied and work well for HHG in gases. Several particle-like recollision models for HHG in solids have also been developed and used to explore HHG features such as the emission time, harmonic cutoff, etc. \cite{vampa2015semiclassical,osika2017,yue2020,parks2020}. Despite being promising in some cases, our discussions above suggest that there are still challenges for the recollision picture to describe solid HHG considering the non-parabolic energy band and the delocalization of the valence electrons for general semiconductors, insulators, and dielectrics, such as the noticeable error in describing the emission time. This prevents us from extending the attosecond technologies from gases to solid systems.
	
The success and challenge of the particle-like recollision picture based on the saddle-point approximation are reminiscent of the use of geometric optics. Although the propagation of light should be accurately described as waves, one may approximately describe it in form of rays, in analogy with the semiclassical trajectories in HHG, see Fig. \ref{vs_optics}. In fact, the geometric-optics approximation (or eikonal approximation) is just one of the applications of the saddle point approximation. However, when the size of objects and apertures is comparable to the wavelength of light, geometry optics is no more a good approximation. Diffraction effects become noticeable, which are essentially interference of wavelets, and one must restore to wave optics described by the Huygens-Fresnel principle. Likewise, when saddle-point approximation is not good enough, one should restore to more accurate models taking into account the wave features of electron wavepackets during the HHG process. In the following section, we will introduce such kind of model in the spirit of the Huygens-Fresnel principle.

\section{3. Huygens-Fresnel picture for HHG in solids}
In this section, we introduce a wave perspective of solid HHG by an analogy to the Huygens-Fresnel principle. Aiming to show the mathematical and physical basis of the wave perspective more intuitively, we use the propagator to re-deduce the interband current in a form of Feynman path integral. There is no essential difference from the derivations shown in Sec. 2.A, provided the same approximations are applied.

\begin{figure}[!t]
	\includegraphics[width=\columnwidth]{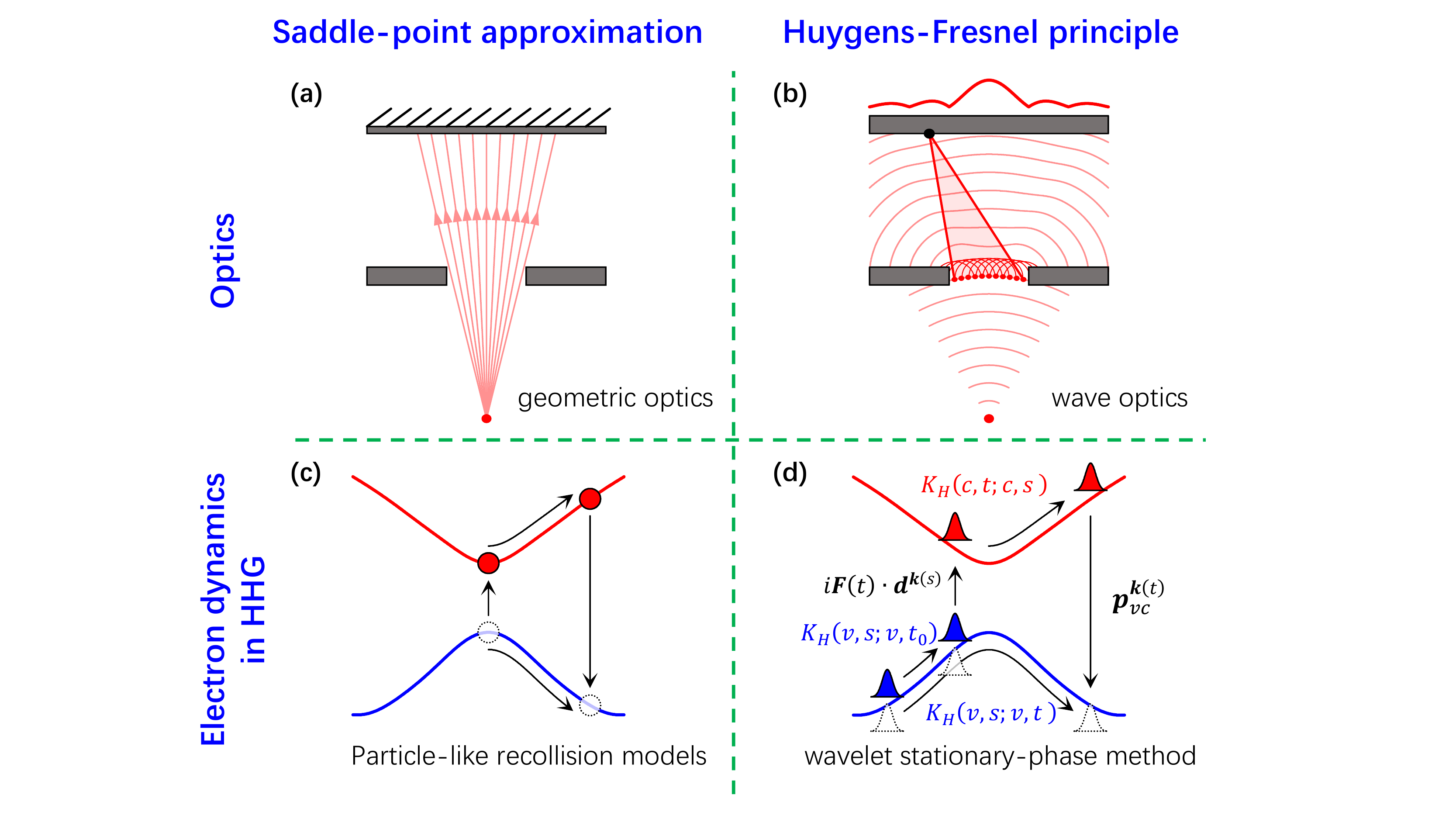}
	\caption{Relation between the particle-like recollision models and wacelet stationary-phase method.}\label{vs_optics}
\end{figure}

A general propagator $\hat{K}(t,t_{0})$, describing the propagation of a state from $t_{0}$ to $t$, satisfies fundamental properties as
\begin{align}\label{Propagator}
|\Psi(t)\rangle &= \hat{K}(t,t_{0})|\Psi(t_{0})\rangle,   \\
\hat{K}(t,t_{0}) &= \hat{K}(t,t')\hat{K}(t',t_{0}).
\end{align}
Using the propagator, one can rewrite the current contributed by an electron with an initial momentum $\textbf{k}_{0}$ as
\begin{align}\label{interband current using propagator}
\textbf{j}_{\textbf{k}_{0}}(t) &= -\langle\varphi_{v,\textbf{k}_{0}}(t_{0})|\hat{K}_{H}(t_{0},t)\left[\hat{\textbf{p}}+\textbf{A}(t)\right]\hat{K}_{H}(t,t_{0})|\varphi_{v,\textbf{k}_{0}}{(t_{0})}\rangle,   \nonumber\\
&= -\sum_{m,n}\langle\varphi_{v,\textbf{k}_{0}}(t_{0})|\hat{K}_{H}(t_{0},t)|\varphi_{m,\textbf{k}_{0}}(t)\rangle\textbf{p}_{mn}^{\textbf{k}_{0}+\textbf{A}(t)}\langle\varphi_{n,\textbf{k}_{0}}(t)|\hat{K}_{H}(t,t_{0})|\varphi_{v,\textbf{k}_{0}}(t_{0})\rangle.
\end{align}
In this form, we transform the expression to Houston representation, i.e., using the propagator $\hat{K}_{H}(t,t_{0}) = \mathcal{T}e^{-i\int_{t_{0}}^{t}\tilde{H}(s)\mathrm{d}s}$ with a system Hamiltonian $\tilde{H}_{mn}(t) = \delta_{mn}E_{n}\textbf{(}\textbf{k}(t)\textbf{)}+\textbf{F}(t)\cdot\textbf{d}_{mn}\textbf{(}\textbf{k}(t)\textbf{)}$. Then, by using the same approximation as in Sec. 2.B, the interband current in a two-band model can be expressed in the form
\begin{align}\label{Feynman path integral form interband current}
\textbf{j}_{\textbf{k}_{0}}^{\text{ter}}(t) &= -\langle\varphi_{v,\textbf{k}_{0}}(t_{0})|\hat{K}_{H}(t_{0},t)|\varphi_{v,\textbf{k}_{0}}(t)\rangle\textbf{p}_{vc}^{\textbf{k}_{0}+\textbf{A}(t)}\langle\varphi_{c,\textbf{k}_{0}}(t)|\hat{K}_{H}(t,t_{0})|\varphi_{v,\textbf{k}_{0}}(t_{0})\rangle + c.c.,   \nonumber\\
&= -\int_{s}^{t}\mathrm{d}sK_{H}(v,t_{0};v,t)\textbf{p}_{vc}^{\textbf{k}(t)}K_{H}(c,t;c,s)\left[-i\textbf{F}(t)\cdot\textbf{d}^{\textbf{k}(s)}\right]K_{H}(v,s;v,t_{0}) + c.c.,
\end{align}
where $K_{H}(m,t_{2};m,t_{1}) = e^{-i\int_{t_{1}}^{t_{2}}\left[E_{m}\textbf{(}\textbf{k}(t')\textbf{)}+F(t')\cdot\mathbf{\Lambda}_{c}^{\textbf{k}(t')}\right]\mathrm{d}t'}$ describes the intraband propagation from time $t_{1}$ to $t_{2}$ on band $m$. The second line in Eq. (\ref{Feynman path integral form interband current}) is equivalent to Eq. (35). Moreover it suggests a Feynman path interpretation of interband HHG as illustrated in Fig. \ref{vs_optics}(d): (i) an electron initially ($t_{0}$) located at $\textbf{k}_{0}$ in the valence band is propagated to $\textbf{k}(s)$ at time $s$, and (ii) is ionized from valence to conduction band; (iii) the ionized electron is propagated from time $s$ to $t$ in the conduction band, and (iv) recombine with the hole propagating from time $t_{0}$ to $t$ in the valence band. This is exactly the physical picture the four-step model depicts \cite{Li_2019}, and the pre-acceleration process is none other than the electron propagation described by the term $K_{H}(v,s;v,t_{0})$ before ionization. The pre-acceleration process suggests that one can selectively excite the electron from different initial crystal momentum dominantly contributing to HHG by designing the laser fields, so as to realize the control of HHG (yield, cutoff energy, etc.), and to access electronic and optical properties for materials in a wider region of BZ by using HHS.

As has been discussed in Sec. 2.B, we have shown that the electron wave packet will be dramatically distorted during the evolution due to the delocalization of the electron wave packet and the complicated dispersion of the band structure in solids. Therefore, the wave properties of electrons have to be considered. Here, a wave perspective of solid HHG by an analogy to the Huygens-Fresnel principle is introduced \cite{Li_2021}: the electron wave packet ionized by the laser field is treated as a composition of wavelets in analogy with the secondary wavelets in the Huygens-Fresnel principle. Each wavelet is denoted as $\{\textbf{k}_{l},s\}$, where $\textbf{k}_{l}$ is the central momentum of the electron wavelet at time $s$. In this case, observing the harmonic emission at time $t_{r}$, just like observing the light wave at a given observation point in the Huygens-Fresnel principle, can be described by the interference of the contributions from all wavelets. Following these concepts, one can express the harmonic emission at time $t_{r}$ as:
\begin{align}\label{Huygens-Fresnel formed interband current}
Y_{\mu}(t_{r},\omega) = \sum_{\textbf{k}_{l},s}f(\textbf{k}_{l},s)\int_{-\infty}^{+\infty}\int_{-\infty}^{+\infty}\mathrm{d}t\mathrm{d}\textbf{k}'g(\textbf{k}',\textbf{k}_{l})e^{-iS(\textbf{k}',t,s)}R(\textbf{k}',t,s)w(t,t_{r})+c.c.,
\end{align}
where $w(t,t_{r}) = e^{-\left(t-t_{r}\right)^{2}/t_{w}^{2}}$ is an integral window with width $t_{w}$ near the observation point. In this form, we separate the electron wave packet into a series of Gaussian wavelets $g(\textbf{k}',\textbf{k}_{l}) = e^{-\left(\textbf{k}'-\textbf{k}_{l}\right)^{2}/k_{w}^{2}}$ with width $k_{w}$. The corresponding weight coefficient $f(\textbf{k}_{l},s)$ satisfies $\sum_{k_{l}}g(\textbf{k}',\textbf{k}_{l})f(\textbf{k}_{l},s) = T^{\textbf{k}',s} = F(s)\left|\textbf{d}_{\|}^{\textbf{k}'}\right|$. The accumulated phase $S(\textbf{k}',t,s)$ and the disturbance $R(\textbf{k}',t,s)$ of the wavelet $\{\textbf{k}_{l},s\}$ are
\begin{align}\label{phase and disturbance}
S(\textbf{k}',t,s) &= -\text{arg}\left[K_{H}(v,t_{0};v,t)K_{H}(c,t;c,s)K_{H}(v,s;v,t_{0})\right]+\beta_{\mu}^{\textbf{k}(\textbf{k}',t,s)}-\beta_{\|}^{\textbf{k}'}-\omega t  \nonumber\\
&= \int_{s}^{t}\left[\omega_{g}^{\textbf{k}(\textbf{k}',t',s)}+\textbf{F}(t')\cdot\Lambda_{cv}^{\textbf{k}(\textbf{k}',t',s)}\right]\mathrm{d}t'+\beta_{\mu}^{\textbf{k}(\textbf{k}',t,s)}-\beta_{\|}^{\textbf{k}'}-\omega t,   \\
R(\textbf{k}',t,s) &= \omega_{g}^{\textbf{k}(\textbf{k}',t,s)}\left|\textbf{d}_{\mu}^{\textbf{k}(\textbf{k}',t,s)}\right|, 
\end{align}
with $\textbf{k}(\textbf{k}',t,s) = \textbf{k}'+\textbf{A}(t)-\textbf{A}(s)$.

For brevity, we first focus on the contribution of a single wavelet
\begin{align}\label{single wavelet contribution}
D(k_{l},t_{r},s) = \int_{-\infty}^{+\infty}\int_{-\infty}^{+\infty}\mathrm{d}t\mathrm{d}\textbf{k}'g(\textbf{k}',\textbf{k}_{l})e^{-iS(\textbf{k}',t,s)}R(\textbf{k}',t,s)w(t,t_{r}).
\end{align}
Here, we use a narrow time window to probe the harmonic emission near the observation time $t_{r}$, so that one can apply the first order Taylor expansion to the accumulated phase
\begin{align}\label{single wavelet contribution-1}
D(k_{l},t_{r},s) &= \int_{-\infty}^{+\infty}\int_{-\infty}^{+\infty}dtd\textbf{k}'g(\textbf{k}',\textbf{k}_{l})w(t,t_{r})e^{-i\left[S(\textbf{k}',t_{r},s)+S^{(1)}_{t}(\textbf{k}',t_{r},s)(t-t_{r})\right]}R(\textbf{k}',t,s),   \nonumber   \\
& = \int_{-\infty}^{+\infty}d\textbf{k}'g(\textbf{k}',\textbf{k}_{l})e^{-iS(\textbf{k}',t_{r},s)}\int_{-\infty}^{+\infty}dtw(t,t_{r})e^{-iS^{(1)}_{t}(\textbf{k}',t_{r},s)(t-t_{r})}R(\textbf{k}',t,s),   \nonumber   \\
& = \int_{-\infty}^{+\infty}d\textbf{k}'g(\textbf{k}',\textbf{k}_{l})e^{-iS(\textbf{k}',t_{r},s)}\int_{-\infty}^{+\infty}d\tau w(\tau,0)e^{-iS^{(1)}_{t}(\textbf{k}',t_{r},s)\tau}R(\textbf{k}',t_{r}+\tau,s),   \nonumber\\
& = \int_{-\infty}^{+\infty}d\textbf{k}'g(\textbf{k}',\textbf{k}_{l})e^{-iS(\textbf{k}',t_{r},s)}R(\textbf{k}',t_{r},s)W\left[S^{(1)}_{t}(\textbf{k}',t_{r},s),\Delta E\right].
\end{align}
In the third line, we have replaced the integration variable $t$ with $\tau = t-t_{r}$, and the last line comes from the integration over $\tau$ using Gaussian integration, where we use the notation
\begin{align}\label{Gaussian contribution for t}
W\left[S^{(1)}_{t}(\textbf{k}',t_{r},s),\Delta E\right] = \int_{-\infty}^{+\infty}d\tau w(\tau,0)e^{-iS^{(1)}_{t}(\textbf{k}',t_{r},s)\tau} = t_{w}\sqrt{\pi}e^{-\left[\frac{S^{(1)}_{t}\left(\textbf{k}',t_{r},s\right)}{\Delta E}\right]^{2}}.
\end{align}
$S^{(1)}_{t}$ denotes the partial derivative with respective to $t$ and $\Delta E = 2/t_{w}$ is the emission bandwidth. The integral over $\textbf{k}'$ can also be performed following similar procedure and Eq. (\ref{single wavelet contribution-1}) can be further derived as
\begin{align}\label{single wavelet contribution-2}
D(\textbf{k}',t_{r},s) &= e^{-iS(\textbf{k}_{l},t_{r},s)}G\left[S^{(1)}_{\textbf{k}'}(\textbf{k}_{l},t_{r},s),\Delta x\right]W\left[S^{(1)}_{t}(\textbf{k}_{l},t_{r},s),\Delta E\right]R(\textbf{k}_{l},t_{r},s)
\end{align}
where $G\left[S^{(1)}_{\textbf{k}'}(\textbf{k}_{l},t_{r},s),\Delta x\right] = (k_{w}\sqrt{\pi})^{3}e^{-\left[\frac{S^{(1)}_{\textbf{k}'}(\textbf{k}_{l},t_{r},s)}{\Delta x}\right]^{2}}$ with $\Delta x = 2/k_{w}$ and $S^{(1)}_{\textbf{k}'}(\textbf{k}_{l},t_{r},s) = \nabla_{\textbf{k}'}S(\textbf{k}',t_{r},s)|_{\textbf{k}' = \textbf{k}_{l}}$.

By substituting Eq. (\ref{single wavelet contribution-2}) into Eq. (\ref{Huygens-Fresnel formed interband current}), the harmonic yield can be rewritten as
\begin{align}\label{harmoic emission in Huygens-Fresnel picture}
Y_{\mu}(t_{r},\omega) = \sum_{\textbf{k}_{l},s}f(\textbf{k}_{l},s)P(\textbf{k}_{l},t_{r},s)e^{-iS(\textbf{k}_{l},t_{r},s)}
\end{align}
where $P(\textbf{k}_{l},t_{r},s) = G\left[S^{(1)}_{\textbf{k}'}(\textbf{k}_{l},t_{r},s),\Delta x\right]W\left[S^{(1)}_{t}(\textbf{k}_{l},t_{r},s),\Delta E\right]R(\textbf{k}_{l},t_{r},s)$ is
a Gaussian form emission pulse contributed by a single wavelet $\{\textbf{k}_{l},s\}$ and $e^{-iS(\textbf{k}_{l},t_{r},s)}$ is the corresponding phase. Equation (\ref{harmoic emission in Huygens-Fresnel picture}) can be intuitively interpreted in form of Huygens-Fresnel principle: secondary wavelets from the source with weights $f(\textbf{k}_{l},s)$ coherently disturb the observation point with amplitude $P(\textbf{k}_{l},t_{r},s)$ and relative phases $e^{-iS(\textbf{k}_{l},t_{r},s)}$.

\section{4. Comparison between wave and particle perspective}
From a traditional particle perspective, one can assume that a harmonic emission with precise photon energy is emitted at a certain time. From a wave perspective, the dominant contribution of a single wavelet is determined by the conditions
\begin{align}\label{pulse central}
\left|S^{(1)}_{t}(\textbf{k}_{l},t_{r},s)\right| &=  \left|\omega_{g}^{\textbf{k}_{l}+\textbf{A}(t_{r})-\textbf{A}(s)}+\textbf{F}(t_{r})\cdot\textbf{D}_{\mu}^{\textbf{k}_{l}+\textbf{A}(t_{r})-\textbf{A}(s)}-\omega\right| < \Delta E,   \\
\left|S^{(1)}_{k'}(\textbf{k}_{l},t_{r},s)\right| &= \left|\Delta\textbf{r}(\textbf{k}_{l},t_{r},s)-\Delta\textbf{D}(\textbf{k}_{l},t_{r},s)\right|<\Delta x.
\end{align}
This means the harmonic emissions at the observation time $t_r$ have uncertainties $\Delta E$ and $\Delta x$ considering the width of the Gaussian distribution $P(\textbf{k}_{l},t_{r},s)$. The wave properties are inherently embedded in the evolution of the Gaussian wavelets, and the contributions by wavelets at different $t_{r}$ can be fully interfered. By contrast, in the particle perspective, the saddle point contribution with different emission times and frequencies are independent with each other. 

For the summation over $s$, the constructive interference occurs when the phase varies most slowly, i.e.
\begin{align}\label{constructive ionization time}
&\left|\partial_{s}S(\textbf{k}_{l},t_{r},s)\right| = \min\left\{\left|\partial_{s}S(\textbf{k}_{l},t_{r},s)\right|\right\},   \nonumber\\
\Leftrightarrow\ & \left|-\omega_{g}^{\textbf{k}_{l}}+ \textbf{F}(s)\cdot\left[\Delta\textbf{r}(\textbf{k}_{l},t_{r},s)-\textbf{D}_{\mu}^{\textbf{k}_{l}+\textbf{A}(t_{r})-\textbf{A}(s)}\right]\right| = \min\left\{\left|\partial_{s}S(\textbf{k}_{l},t_{r},s)\right|\right\}.
\end{align}
These conditions are different from that obtained by saddle point equations. The classical correspondence between the microscopic electron dynamics and harmonic emissions is established by considering the most probable distribution at the observation point. Note that particle-like recollision picture from the saddle point method can be reproduced in a quasiparticle limit, i.e., $\Delta x\rightarrow 0$ and $\Delta E\rightarrow 0$. This right corresponds to the limitation where the integrand phase term varies much faster than the amplitude factors. In this case, the saddle point method gives a considerable approximation. Therefore, the wave perspective indeed involves the particle-like recollision pictures as a subset case, and gives more comprehensive physical insights. More specifically, we can take the simple symmetric one-dimensional system as an example to show the wave properties. In this case, the condition (\ref{constructive ionization time}) can be deduced to $\Delta r(\textbf{k}_{l},t_{r},s) = -\frac{F(s)^{2}}{\partial_{s}F(s)}\int_{s}^{t_{r}}\partial_{k}^{2}\omega_{g}^{k(k_{l},t',s)}dt'$ (for $\min\left\{\left|\partial_{s}S(\textbf{k}_{l},t_{r},s)\right|\right\}>0$). The term $\int_{s}^{t_{r}}\partial_{k}^{2}\omega_{g}^{k(k_{l},t',s)}dt'$ just corresponds to the wave deformation during propagation with the dispersion $\omega_{g}^{k(k_{l},t',s)}$, which evaluates the influence of the wave properties.

\begin{figure}[!t]
	\includegraphics[width=17cm]{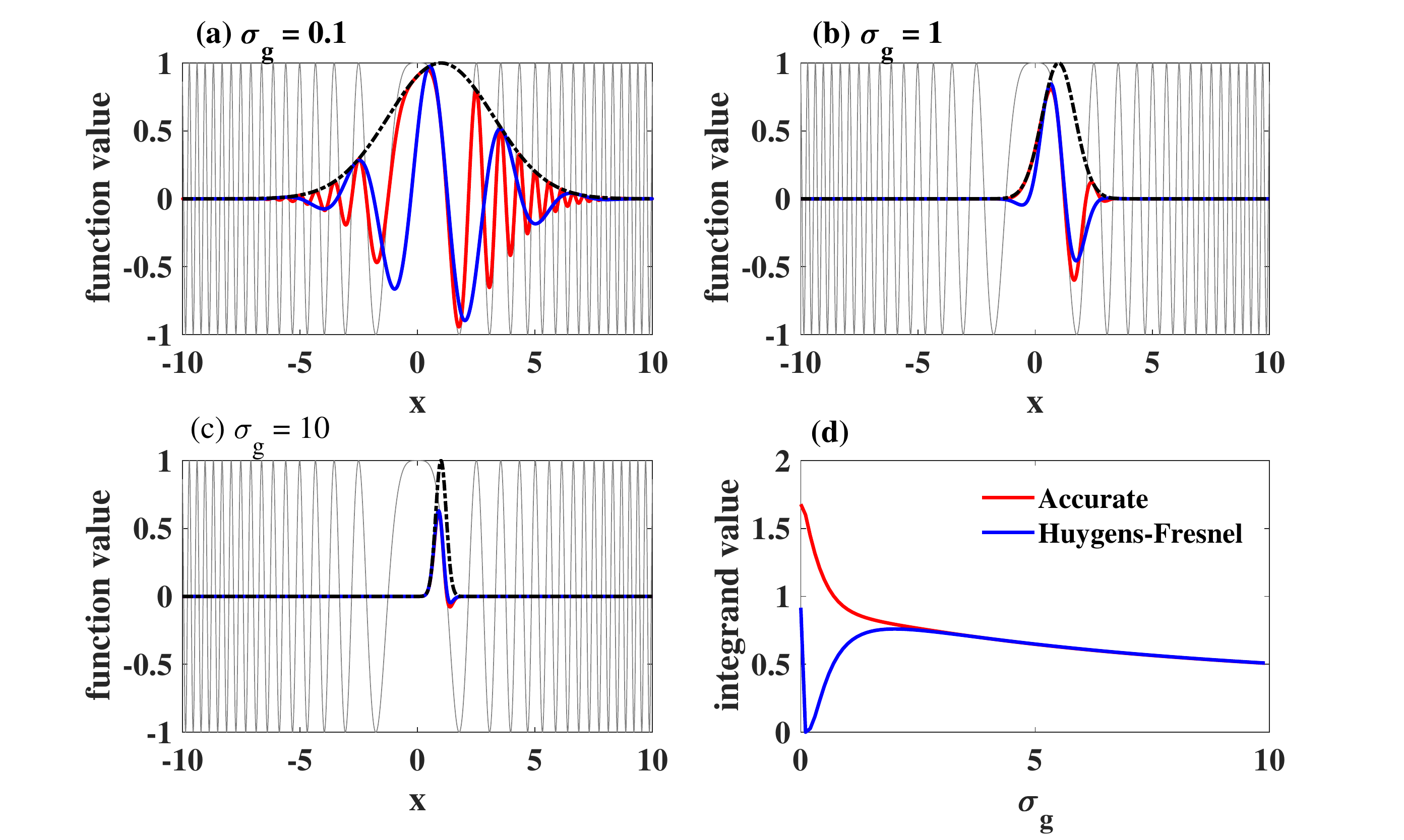}
	\caption{The behavior of Huygens-Fresnel method for $f(z) = e^{-\sigma_{g}(z-1)^{2}}$ and $g(z) = e^{iz^{2}}$. (a)-(c) The gray and dashed black lines denote the function value of $f(z)$ and $\text{Re}(g(z))$ along the x-axis. The integrand function values for accurate and Huygens-Fresnel methods are shown as red and blue lines for a wavelet with (a) $\sigma_{g} = 0.1$, (b) $\sigma_{g} = 1$, and (c) $\sigma_{g} = 10$. (d) The comparison between the integrand value for accurate and Huygens-Fresnel methods for different parameter $\sigma_{g}$.}\label{HF-behavior}
\end{figure}

Having established the Huygens-Fresnel model, we continue the discussions in Sec. 2.C, so that one can intuitively illustrate the connections and differences between the Huygens-Fresnel and saddle point methods. Here, we still focus on the oscillatory integration $I(\sigma) = \int_{-\infty}^{+\infty}f(z)e^{\sigma g(z)}\mathrm{d}z$. Different from the saddle point approximation, which is only valid for large $\sigma$, Huygens-Fresnel method is realized based on a wavelet ensemble and can give reasonable results provided the Gaussian wavelets are narrow enough. Thus, the accuracy of the Huygens-Fresnel method is controllable and can be always guaranteed. To the contrary, the validation of the saddle point method relies on the system parameters, e.g., the electron band structure and the laser parameter, and may fail in some circumstances (as has been shown in Sec. 2.C). 

This does not mean that the saddle-point method is not valuable. It can still be used to simplify the Huygens-Fresnel calculation and facilitate our analysis on the HHG process even for $\sigma\sim1$, as one needs to only take into account the wavelets near the saddle points to get good accuracy. For an intuitive demonstration, we come to the specific functional form $I(\sigma) = \int_{-\infty}^{+\infty}e^{\sigma_{g}(z-1)^{2}}e^{iz^{2}}\mathrm{d}x$. The behavior of the integrand for different $\sigma_g$ is shown in Figs. \ref{HF-behavior}(a)-(c).  The integration is performed along the real axis, which directly corresponds to physical quantity without introducing the complex part. Along this path, the oscillatory term varies faster when leaving further away from the saddle point (see the gray lines in Fig. \ref{HF-behavior}). This indicates that the dominant contribution still comes from the region near the saddle points. Thus, only the contribution of wavelets near the saddle points needs to be considered.  The contribution of a single wavelet from the Huygens-Fresnel method ($\text{Re}(e^{\sigma_{g}(x-1)^{2}}e^{i+i2(x-1)})$) is also shown in Figs. \ref{HF-behavior}(a)-(c) for comparision.  One can see that the discrepancy between the accurate integrand function and the contribution of a single Huygens-Fresnel wavelet becomes smaller for large $\sigma_g$, where the width of the integrand function (with a Gaussian envelope $f(z) = e^{-\sigma_{g}(z-1)^{2}}$) is smaller. This behavior is further demonstrated in Figures \ref{HF-behavior}(d) by comparing the integration versus width parameter $\sigma_{g}$. With increasing $\sigma_{g}$ (i.e., narrower wavelet), the result of Huygens-Fresnel method converges rapidly to the accurate one.

The relation between the particle perspective recollision and wave perspective Huygens-Fresnel pictures is indeed analogous to the different limits of a single slit diffraction. For a slit much wider than the wavelength, the phenomenon of geometric optics will cover up the diffraction phenomenon. Namely, the influence of the latter is very small compared with the geometric optical phenomenon dominated by Fermat's principle even though the diffraction phenomenon still exists under the influence of the boundary. However, the wave properties are more and more significant as the slit width decreases and become comparable to the wavelength. The diffraction pattern should be described by the interference of the wavelets following the Huygens-Fresnel principle. As expected, the integral for HHG near the saddle point has a similar form to the Fresnel-Kirchhoff diffraction integral. The wave perspective Huygens-Fresnel picture for HHG is none other than a counterpart of the optical Huygens-Fresnel principle. Meanwhile, the particle perspective recollision picture of HHG can be seen as a counterpart of the geometric-optics approximation by treating the propagation of light as rays. 

In the above discussions, we mainly focus on the mathematical and physical bases of the Huygens-Fresnel and recollision pictures. For a semi-classical model, it is expected to provide a simple and intuitive physical picture and help people understand the dynamical mechanism behind the observations. Both Huygens-Fresnel and recollision pictures provide intuitive physical insights in the HHG process. However, their performances in specific processes are different. In the recollision picture, an initial tunneling time $s$ and an initial crystal momentum $k_{0}$ are picked, and the emission time is determined by the condition of recollision, i.e., $\nabla_{\textbf{k}}S^{\mu}(\textbf{k},t,s) = \Delta \textbf{r}(\textbf{k},t,s)-\Delta \textbf{D}(\textbf{k},t,s) = 0$. In contrast, in the Huygens-Fresnel picture, the semiclassical correspondence is established on the constructive interference condition of different wavelets $\left|\partial_{s}S(\textbf{k}_{l},t_{r},s)\right| = \min\left\{\left|\partial_{s}S(\textbf{k}_{l},t_{r},s)\right|\right\}$, i.e., for an observation point $\{\omega,t_{r}\}$, and one can find the constructive ionization time $s$ for from each wavelet. Note that the Gaussian term $G\left[S^{(1)}_{\textbf{k}'}(\textbf{k}_{l},t_{r},s),\Delta x\right]$ has a large width $\Delta x$ considering the narrow wavelet applied. The significant displacement between the electron and hole $S^{(1)}_{\textbf{k}'}(\textbf{k}_{l},t_{r},s)$ has less important influences on the emission, compared with the picture based on the recollision conditions where the radiation decays rapidly with increasing displacement. This difference, due to the non-local property of the wavepackets in solids, leads to different predictions for both the time-frequency properties and the time-domain interferometry in the HHG process as verified in Ref \cite{li2021}. Moreover, Huygens-Fresnel picture can provide a more comprehensive understanding of the HHG process by analyzing the interference and transition between different wavelets. Specifically, the semi-classical correspondence obtained from the recollision picture is a series of independent paths. The phenomena, such as channel splitting and transforming, have been oversimplified as the saddle point contribution, and the wave properties are not properly involved. In general, the recollision picture can well describe the HHG process under conditions where the saddle point approximation is applicable, as it does in gas HHG. When the saddle-point approximation becomes inadequate, although the recollision picture can still provide qualitative understanding, the Huygens-Fresnel involving the wave property is indispensable, especially for cases an exact and comprehensive understanding of the HHG process is needed, e.g., ultra-high resolution measurements using HHS.

\section{5. Summary and outlook}
In this review, we have given an introduction to the physical pictures of the interband HHG in solids from both particle and wave perspectives. We discuss in detail the mathematical basis, physical interpretation and possible shortcomings of existing recollision models from particle perspective. Then, a Huygens-Fresnel picture from wave perspective is also introduced. The similarities and differences between this model and the recollision models are discussed to illustrate how it overcomes the shortcomings of the existing recollision methods. To some extent, the saddle-point approximation is a mathematical bridge between the quantum world and classical physics. However, one needs to be very careful about the conditions under which it holds. The Huygens-Fresnel picture gives a comprehensive physical insight in the strong-field ultrafast science, and extends the range of application to the region out of reach of the recollision picture. Although in this review we have focused our discussion on the physical picture of HHG in solids, the conclusions we have drawn clearly hold for a broader variety of scenarios:\\
\textbullet\textbf{Visions with HHS in solids}. One can develop attosecond spectroscopy in the condensed matter if an internal clock capable of resolving ultrafast dynamics is provided. Specifically, the interband HHG in solids forms a time domain interferometry, which gives a self-detection of both the crystal structure and the microscopic dynamics \cite{vampa2015,Vampa_2015,Luu_2018,banks2017,hohenleutner2015,jiang2018,Li_2020}. The fact, that the timeline provided by the particle-like picture shows clear deviations, strongly indicates that a corrected internal clock should be established taking into account the wave-like behaviors of electron wavepackets in solid HHG. The Huygens-Fresnel method suggests a new paradigm for HHS, where the qualitative preliminary insight can be obtained with saddle point results and richer and exact insight is quired from the fine ``diffraction fringe'' considering the wave properties.
This suggests that developing specific ``diffraction theories'' for varied parameter intervals may be a treasure trove to be investigated in the future.\\
\textbullet\textbf{Visions with HHG in gases}. Although the recollision picture has achieved great success in gas HHG \cite{Corkum_2007,Calegari_2016,Villeneuve_2018}, previous studies have mainly focused on high energy harmonic emission in the plateau or cutoff region. In recent years, a wealth of experimental phenomena have been observed in the near- and below-threshold region \cite{Burnett_1995,Soifer_2010,Power_2010,Ferr__2014}, which are closely related to the dynamics of low energy electrons and fine orbital structures. Numerical or semi-empirical theories have been developed to understand these phenomena \cite{Kakehata_1997,Wang_2021,Long_2023}. However, the extension of the semiclassical trajectory model is difficult and controversial. The main difficulty is that in these regions the Coulomb effect is no longer negligible. Although theoretical models try to solve this problem by modifying the semiclassical action including the Coulomb potential, the quantum response under the Coulomb potential is not well described. For example, in such models, describing the potential effect in semiclassical action $V(x(t))$ needs to determine both the momentum and location of the electron, which violates the uncertainty principle. In this case, coming back to the scenery of electron wave propagation is favorable. It is easy to find that for a narrow spatial wavepacket, the phase accumulated due to the Coulomb effect can be determined by the path of its central position. Therefore, one can try to decompose the wavepacket propagation under the Coulomb potential into the propagation of a series of spatial wavelets using the Huygens-Fresnel method. This may help to build a uniform semiclassical picture for HHG involving both high and low-energy electrons.\\
\textbullet\textbf{Visions with strong field ionization}. Quantum trajectory models are widely used in strong field ionization, either to interpret the interference structure in the photoelectron momentum spectrum or to construct the LIED \cite{salieres2001,le2009,Li_2014,Milo_evi__2016,Liu_2017,Shvetsov_Shilovski_2019}. In these models, the microscopic dynamics are always described by a trajectory ensemble with semiclassical action determined by the energy accumulation along the classical electron trajectories. One may expect the Coulomb effect becomes more important when the electron ionization position is not far from the nucleus and the ionization momentum distribution is not narrow enough, e.g., in the multiphoton ionization regime or for the rescattering arm of photoelectron holography \cite{Maxwell_2017,Maxwell_2018}. In these regions, the classical drift of the electron is not sufficient to pull the electron out of the Coulomb region quickly enough that a significant diffraction effect will occur due to the combined effect of Coulomb attraction and free diffusion of the electron wavepacket. As a result, the quantum trajectory model tends to be invalid because it underestimates the interference between different trajectories --- the Coulomb effect will obviously lead to coupling between trajectories of different momenta. The Huygens-Fresnel method provides a new perspective, may hopefully broaden the application of photoelectron momentum spectroscopy/LIED, and shed light on controversial issues, e.g., the ionization time.

Strong field physics is a rapidly developing field. Its territory is also expanding rapidly, regarding both the involved laser parameters and species of materials. Possibly because of the big success of the particle-like recollision model in gas HHG, it is always easily extended to a new field and applications without strict thinking. The purpose of this review is to point out a problem which is often neglected in the process of expanding the application range of the present particle-like recollision theories and relevant technologies, that is, whether the original approximation or picture still works well. We hope that this review gives newcomers to get an overview of the topic and provides guidance on the potential development of theories and applications.

We acknowledge fundings from National Natural Science Foundation of China (NSFC) (No. 91950202, No.11934006 , No. 12104172, No. 12225406, No. 12021004).

\bibliography{ref.bib}

\end{document}